\documentclass[reprint,superscriptaddress,amsmath,amssymb,aps]{revtex4-2}

\usepackage{graphicx}
\usepackage[caption=false]{subfig}
\usepackage{multirow}
\usepackage{dcolumn}
\usepackage{bm}
\usepackage{hyperref}
\hypersetup{colorlinks=True,
            linkcolor=blue,
           anchorcolor = blue,
            citecolor = blue,
            filecolor = blue,
            urlcolor = blue }       

\begin{document}
\title{Squeezing enhanced homodyne weak force sensing in cavity optomechanics}
\author{Madan Mohan Mahana}
\email{m.madan@iitg.ac.in}
\affiliation{Department of Physics, Indian Institute of Technology Guwahati, Guwahati 781039, Assam, India}
\author{Sanket Das}
\email{sanket.das@oist.jp}
\affiliation{Quantum Machines Unit, Okinawa Institute of Science and Technology Graduate University, Okinawa 904-0495, Japan}
\author{Tarak Nath Dey}
\email{tarak.dey@iitg.ac.in}
\affiliation{Department of Physics, Indian Institute of Technology Guwahati, Guwahati 781039, Assam, India}
 
\begin{abstract}
Cavity optomechanical systems have emerged as a promising platform for quantum sensing. Quantum mechanics imposes a standard quantum limit (SQL) on the force-sensitivity for the standard homodyne phase quadrature measurement of the cavity's output field. In this paper, we investigate ways to enhance weak force sensitivity beyond SQL by employing a variational homodyne quadrature readout and quantum squeezing. Our study reveals a remarkable improvement in the force sensitivity of a cavity optomechanical sensor at a suitable homodyne angle, compared with standard phase quadrature detection of the cavity output field within a specific frequency band. We further demonstrate improved force sensitivity via intra-cavity squeezing (ICS) or injected external squeezing (IES) of the cavity mode. Both variational homodyne readout and quantum squeezing induce quantum correlations between the amplitude and phase quadratures of the cavity's output field, thereby improving force sensitivity. Our results suggest that IES is preferable to ICS for sub-SQL force sensing with system stability and lower probe power requirements. The squeezing-enhanced variational homodyne detection scheme can enable high-precision quantum sensing across various hybrid quantum platforms.
\end{abstract}
\maketitle
\section{\label{sec:level1}Introduction}
Cavity optomechanics is a notable platform for modern quantum technological applications \cite{RevModPhys.86.1391, Barzanjeh2022}. In a cavity optomechanical system, a mechanical object is coupled to the electromagnetic or optical mode of a cavity via radiation pressure. Advances in manufacturing and fabrication techniques have enabled the embedding of nano- or microscale mechanical resonators in optical cavities or superconducting microwave resonators for the study of cavity optomechanics in the quantum regime \cite{10.1063/5.0021088}. Over the last two decades, cavity quantum optomechanics has advanced rapidly and emerged as a versatile platform for exploring macroscopic quantum phenomena. Cavity optomechanical systems have a wide range of applications, including ground-state cooling of a mechanical oscillator \cite{PhysRevLett.99.093901, PhysRevLett.99.093902, Teufel2011, Chan2011, PhysRevA.111.053505}, quantum sensing \cite{LiOuLeiLiu+2021+2799+2832, 10.1063/5.0237048}, optomechanically induced transparency (OMIT) \cite{doi:10.1126/science.1195596, PhysRevA.81.041803}, quantum memory \cite{PhysRevLett.107.133601, Palomaki2013, PhysRevLett.132.100802, Bozkurt2025}, and gravitational-wave detection \cite{doi:10.1126/science.256.5055.325, PhysRevD.111.062002}. 

Fabricated cavity quantum optomechanical devices can enable more precise control of the detection of mechanical motion in the quantum regime than natural quantum systems, thereby enabling ultrasensitive quantum measurements \cite{Teufel2009, Mason2019}. The sensitivity to mechanical displacement and weak force detection via the standard cavity's output phase quadrature readout is limited by an SQL that optimally balances shot noise from the cavity's input probe field and quantum backaction noise from radiation pressure interaction \cite{PhysRevD.23.1693, Braginsky_Khalili_Thorne_1992}. The shot noise decreases by increasing cavity probe power; conversely, the backaction noise increases. A trade-off between these two produces a minimal added force noise power spectral density at an optimal probe power known as the SQL. There has been an ever-growing interest in achieving sub-SQL sensitivity in cavity optomechanical sensors \cite{PhysRevA.73.033819, PhysRevA.90.043848, PhysRevLett.117.030801}. Introduction of ICS or IES in the cavity \cite{PhysRevLett.115.243603, PhysRevLett.59.278, PhysRevLett.59.2153}, coherent quantum noise cancellation (CQNC) \cite{PhysRevLett.105.123601}, backaction evasion \cite{Clerk_2008, Hertzberg2010}, quantum entanglement \cite{Xia2023}, shot noise-backaction noise quantum correlation \cite{PhysRevD.65.022002}, and feedback-control techniques \cite{PhysRevLett.111.103603, Wilson2015} have been proposed to enhance sensitivity beyond the SQL for mechanical displacement and weak force detection over a wide frequency range. Degenerate parametric amplification of the cavity generates ICS, thereby enhancing the sensitivity of a cavity optomechanical sensor \cite{PhysRevA.95.023844, Zhao2019, PhysRevLett.115.243603, PhysRevA.95.023844}. The CQNC method is based on the principle of complete cancellation of backaction noise by introducing an auxiliary effective negative-mass oscillator as an anti-noise path \cite{PhysRevA.89.053836, PhysRevA.92.043817}. Injecting squeezed vacuum externally introduces IES into the cavity, thereby suppressing the cavity shot noise \cite{9tkg-6xxc, Subhash:23}. A significant enhancement in weak force sensitivity can be achieved by employing CQNC with IES \cite{Motazedifard_2016} or ICS \cite{10.3389/fphy.2023.1142452, PhysRevA.111.013509}. Introducing quantum correlation between the cavity's output amplitude and phase quadratures by placing the local oscillator of a homodyne detector set up at an appropriate homodyne angle is another simple yet powerful approach for sub-SQL quantum sensing in cavity optomechanics \cite{PhysRevX.7.021008, PhysRevLett.121.243601, Mason2019, PhysRevA.110.043512}. 

Motivated by the aforementioned investigations into sub-SQL optomechanical quantum sensing, we employ a squeezing-enhanced variational homodyne detection scheme for a cavity optomechanical weak force sensor \cite{PhysRevX.7.021008}. Our scheme primarily exploits quantum correlations to induce destructive quantum interference between shot noise and backaction noise, thereby enhancing force sensitivity beyond SQL. To observe quantum correlations, a mixture of both the amplitude and phase quadratures of the cavity's output field is read out using a variational homodyne angle. Quantum correlations can increase cavity shot noise. However, within a specific frequency band, they can effectively reduce the total added force noise at certain homodyne angles. Further, introducing ICS or IES into the cavity reduces the added force noise, thereby enabling thermal-noise-limited quantum weak force sensing. We also show that IES outperforms ICS in sub-SQL weak force sensing when combined with a variational homodyne detection scheme. Unlike the CQNC, our scheme does not require hybrid coupling with other quantum systems \cite{PhysRevA.106.023107}.

The paper is organized as follows. Sec. \ref{sec:level2} is devoted to the theoretical description of our proposed model system. In Sec. \ref{sec:level3}, we discuss the variational homodyne weak force detection and its advantage over the standard homodyne phase quadrature readout in the absence of squeezing, with ICS, and with IES in Sec. \ref{sec:level3a}, Sec. \ref{sec:level3b}, and Sec. \ref{sec:level3c}, respectively. Finally, we summarize the key findings of our study in Sec. \ref{sec:level4} and provide a conclusion.
\section{\label{sec:level2}Theoretical Model}
\begin{figure}[t!]
    \centering
    \includegraphics[width=0.45\textwidth]{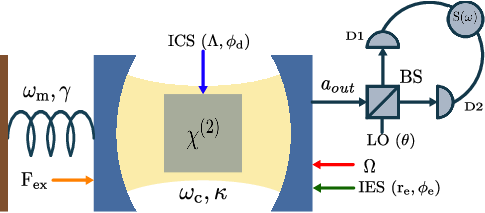}
    \caption{(Color online) The schematic diagram of a cavity optomechanical system with a weak force $F_{ex}$ exerted on the mechanical oscillator. The cavity is pumped by a coherent probe field and a two-photon drive with driving strengths $\Omega$ and $\Lambda$, respectively. The two-photon drive also carries an effective phase $\phi_d$ and can induce ICS in the cavity. The IES refers to the external injection of squeezed vacuum into the cavity with squeezing parameter $r_e$, and squeezing angle $\phi_e$. The cavity output field quadratures are read out via a homodyne detection setup with a $50/50$ beam splitter (BS), detectors ($D1$, $D2$), and a local oscillator (LO) rotated by a homodyne angle $\theta$.}
    \label{fig:fig1}
\end{figure}

We consider an electromechanical system \cite{PhysRevLett.121.243601} consisting of a superconducting microwave cavity resonator with a resonance frequency $\omega_c$ and a mechanical drum oscillator with effective mass $m$ and natural frequency $\omega_m$ as schematically shown in Fig. \ref{fig:fig1}. The decay rates of the cavity and mechanical oscillator (MO) are denoted by the symbols $\kappa$ and $\gamma$, respectively. The resonant mode of the cavity is coupled to the MO via radiation pressure-like interaction $g=-x_\textrm{ZPF}\partial\omega_c/\partial x$, where $x_\textrm{ZPF}=\sqrt{\hbar/(2m\omega_m)}$ is the zero-point fluctuation of the MO's displacement. A probe field coherently drives the microwave cavity mode with a driving frequency $\omega_d$ and power $P$. A degenerate parametric (two-photon) drive with gain $\Lambda$, phase $\phi_d$, and frequency $2\omega_d$ is pumping the cavity. In quantum optics, a $\chi^{(2)}$ nonlinear medium inserted in a single-mode cavity is pumped to induce a squeezed cavity mode \cite{PhysRevA.30.1386, RevModPhys.84.1}. The three-wave-mixing (3WM) type Josephson parametric amplifiers (JPAs), such as flux-pumped superconducting quantum interference devices (SQUIDs) \cite{10.1063/1.2964182}, superconducting nonlinear asymmetric inductive elements (SNAILs) \cite{PhysRevApplied.10.054020}, and kinetic inductance parametric amplifiers (KIPAs) \cite{PhysRevApplied.17.034064} can be used for parametric amplification in a superconducting microwave cavity. A weak external force $F_{ex}$ is exerted on the MO, which is to be detected. The Hamiltonian of the model system can be expressed as
\begin{align}\label{1}
  \hat{H}_0/\hbar=& \omega_c\hat{a}^\dag\hat{a}+\omega_m\hat{b}^\dag\hat{b}-g\hat{a}^\dag\hat{a}(\hat{b}^\dag+\hat{b})-\frac{\hat{F}_{ex}}{\hbar}x_\textrm{ZPF}(\hat{b}+\hat{b}^\dag)\nonumber\\
  &+i\Omega\hat{a}^\dag e^{-i\omega_dt}+i\Lambda\hat{a}^{\dag^2} e^{-i(2\omega_dt+\phi_d)}+H.c.,
\end{align}
where, $\Omega=\sqrt{P\kappa/(\hbar\omega_d)}$ is the driving strength of the coherent probe field. The annihilation and creation operators of the cavity (MO) are denoted by $\hat{a}$ ($\hat{b}$) and $\hat{a}^\dag$ ($\hat{b}^\dag$). The first and second terms in Eq. (\ref{1}) represent the free Hamiltonians of the cavity and the MO, respectively. The third term stands for the optomechanical interaction via radiation pressure. The fourth term describes the weak external force acting on the MO with $\hat{F}_{ex}$ denoting the external force signal. The fifth and sixth terms denote, respectively, the coherent probe with driving strength $\Omega$ and the parametric drive with driving strength $\Lambda$, and phase $\phi_d$. Applying $\hat{U}=e^{i\omega_d\hat{a}^\dag\hat{a}t}$, we obtain a Hamiltonian under the rotating wave approximation (RWA),
\begin{align}\label{2}
 \hat{H}/\hbar=& \hat{U}(\hat{H}_0/\hbar)\hat{U}^\dag-i\hat{U}\frac{\partial\hat{U}^\dag}{\partial t}\nonumber\\
 =& \Delta_c\hat{a}^\dag\hat{a}+\omega_m\hat{b}^\dag\hat{b}-g\hat{a}^\dag\hat{a}(\hat{b}^\dag+\hat{b})-\frac{\hat{F}_{ex}}{\hbar}x_\textrm{ZPF}(\hat{b}+\hat{b}^\dag)\nonumber\\
  &+i\Omega\hat{a}^\dag+i\Lambda\hat{a}^{\dag^2} e^{-i\phi_d}+H.c.,
\end{align}
where $\Delta_c=\omega_c-\omega_d$ is the detuning between the frequencies of the cavity and the external coherent probe field. We can split the operators of the cavity and MO into the steady-state average coherent amplitudes and the fluctuating terms as $\hat{a}=\alpha+\delta\hat{a}$ and $\hat{b}=\beta+\delta\hat{b}$ with $\alpha=\langle\hat{a}\rangle$ and $\beta=\langle\hat{b}\rangle$ by following the linearized approximation of cavity optomechanics for strong driving of the cavity mode (\textit{i.e.,} $\langle\hat{a}^\dag\hat{a}\rangle=|\alpha|^2\gg1$). By keeping the terms in the first order of quantum fluctuations, we can obtain the linearized quantum Langevin equations 
\begin{align}
 \dot{\hat{a}} &= -\left[i\Delta+\frac{\kappa}{2}\right]\hat{a}+iG(\hat{b}+\hat{b}^\dag)+2\Lambda\hat{a}^\dag e^{-i\phi_d}+\sqrt{\kappa}\hat{a}_\textrm{in},\label{3}\\
 \dot{\hat{b}}&=-\left[i\omega_m+\frac{\gamma}{2}\right]\hat{b}+i(G\hat{a}^\dag+G^*\hat{a})+\frac{i\hat{F}_{ex}}{\hbar}x_\textrm{ZPF}+\sqrt{\gamma}\hat{b}_\textrm{in},\label{4}
\end{align}
where we have dropped `$\delta$' from the quantum fluctuations for brevity. Here, $\Delta=\Delta_c-g(\beta+\beta^*)$ is the effective cavity-detuning, $\beta=ig|\alpha|^2/(i\omega_m+\gamma/2)$, $G=g\alpha$ is the linearized optomechanical coupling, and $\alpha=|\alpha|e^{i\psi}$ with 
\begin{subequations}
\begin{align}
|\alpha|=&\frac{\Omega\sqrt{[\{\kappa/2+2\Lambda\cos(\phi_d)\}^2+\{\Delta+2\Lambda\sin(\phi_d)\}^2]}}{[\Delta^2+\kappa^2/4-(2\Lambda)^2]},\label{5a}\\
\psi=&\tan^{-1}\left[\frac{-\{\Delta+2\Lambda\sin(\phi_d)\}}{\kappa/2+2\Lambda\cos(\phi_d)}\right].\label{5b}
\end{align}
\end{subequations}
The operators $\hat{a}_\textrm{in}$ and $\hat{b}_\textrm{in}$ denote the input noise entering the system through the cavity and the MO, respectively. The Eqs. (\ref{3}) and (\ref{4}) in the frequency domain read
 \begin{align}
\frac{\hat{a}(\omega)}{\chi_c(\omega)} &= iG[\hat{b}(\omega)+\hat{b}^\dag(\omega)]+2\Lambda\hat{a}^\dag(\omega) e^{-i\phi_d}+\sqrt{\kappa}\hat{a}_\textrm{in}(\omega),\label{6}\\
\frac{\hat{b}(\omega)}{\chi_m(\omega)}&=i[G\hat{a}^\dag(\omega)+G^*\hat{a}(\omega)]+\frac{i\hat{F}_{ex}(\omega)}{\hbar}x_{ZPF}+\sqrt{\gamma}\hat{b}_\textrm{in}(\omega),\label{7}
\end{align}
where $\hat{\mathcal{O}}(\omega)=\int\limits_{-\infty}^\infty dt e^{i\omega t}\hat{\mathcal{O}}(t)$, ($\hat{\mathcal{O}}=\hat{a}$ or $\hat{b}$). We have also introduced $\chi_c(\omega)=[i(\Delta-\omega)+\kappa/2]^{-1}$ and $\chi_m(\omega)=[i(\omega_m-\omega)+\gamma/2]^{-1}$ as the susceptibilities of the cavity and MO, respectively. Here, 
$\omega$ denotes the Fourier frequency of fluctuations of the input cavity probe field around the driving frequency $\omega_d$. From Eqs. (\ref{6}), (\ref{7}) and Eqs. (\ref{19i}) ,(\ref{19ii}) of Appendix \ref{app:level5}, we can obtain the steady-state solutions of $\hat{a}(\omega)$ and $\hat{a}^\dag(\omega)$. According to the input-output theory for open quantum systems, the field that is reflected from the cavity is given by \cite{114042}
\begin{align}\label{8}
 \hat{a}_\textrm{out}(\omega)=\sqrt{\kappa}\hat{a}(\omega)-\hat{a}_\textrm{in}(\omega).
\end{align}
The exact solution for the cavity's output field for our system is given by Eq. (\ref{22}). 
\section{\label{sec:level3}Homodyne weak force measurement}
\begin{figure*}[t!]
    \centering
    \includegraphics[width=0.98\textwidth]{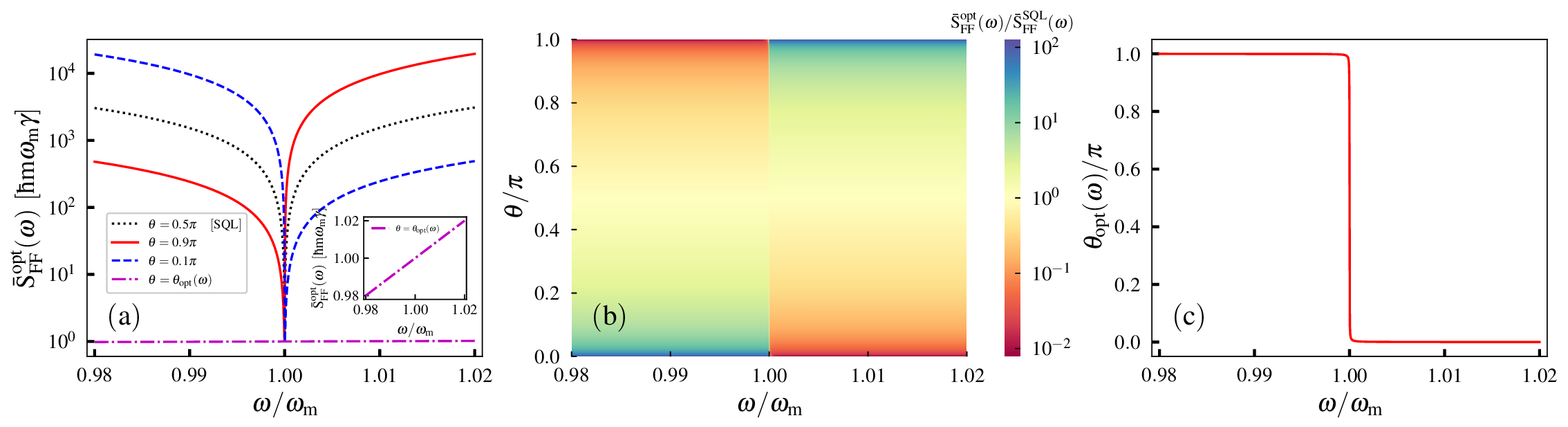}
    \caption{(Color online) (a) The optimal added force noise power spectral density $\bar{S}_{FF}^{opt}(\omega)$ (in the units of $\hbar m\omega_m\gamma$) vs $\omega/\omega_m$ for different values of $\theta$. (b) The surface plot shows the ratio of the optimal added force noise power spectral density to that at SQL as a function of the homodyne angle $\theta$ and  $\omega/\omega_m$. (c) The optimal homodyne angle $\theta_{opt}(\omega)$ vs $\omega/\omega_m$. The parameters used for this figure are $\Delta=0$, $\omega_m/{2\pi}=9.2$ MHz, $\kappa/2\pi=27.8$ MHz, and $\gamma/2\pi=120$ Hz \cite{PhysRevLett.121.243601}. The other parameters are $\omega_c/2\pi=6.4$ GHz, $g/2\pi=260$ Hz, mass of the MO is $m=85$ pg \cite{Clark2017ER, PhysRevLett.116.013602}, and $\hbar m\omega_m\gamma=3.91\times10^{-37}$ $\mathrm{N^2Hz^{-1}}$.}
    \label{fig:fig2}
\end{figure*}
In cavity optomechanics, the optical cavity acts as an interferometer to measure the motion of the MO \cite{10.1063/5.0237048}. The cavity output field carries information about mechanical motion due to the optomechanical radiation-pressure interaction. Linear detection methods, such as homodyne \cite{Teufel2009} or heterodyne \cite{Brooks2012} techniques, are used to extract information about mechanical motion. An external force acting on the MO shifts its position and affects the intra-cavity field $\hat{a}(\omega)$. The signal associated with the external force exerted on the MO can be extracted by measuring the quadratures of the cavity output field. A generalized homodyne quadrature of the cavity output field can be defined as a linear combination of both amplitude $\hat{X}_{a,\textrm{out}}(\omega)=[\hat{a}_\textrm{out}(\omega)+\hat{a}^\dagger_\textrm{out}(\omega)]/\sqrt{2}$ and phase $\hat{P}_{a,\textrm{out}}(\omega)=i[\hat{a}^\dag_{\textrm{out}}(\omega)-\hat{a}_{\textrm{out}}(\omega)]/\sqrt{2}$ as
\begin{align}\label{9}
 \hat{Z}_\textrm{out}(\omega,\theta)=&\hat{X}_{a,\textrm{out}}(\omega)\cos(\theta)+\hat{P}_{a,\textrm{out}}(\omega)\sin(\theta)\nonumber\\
 =&\frac{1}{\sqrt{2}}[\hat{a}_\textrm{out}(\omega)e^{-i\theta}+\hat{a}_\textrm{out}^\dagger(\omega)e^{i\theta}],
\end{align}
where $\theta$ denotes the phase angle of the LO in a homodyne detector setup, called the homodyne angle. Substituting Eqs. (\ref{22}) of Appendix \ref{app:level5} in Eq. (\ref{9}), we obtain
\begin{align}\label{10}
 \hat{Z}_{out}(\omega,\theta)=&\mathcal{G}(\omega,\theta)[\hat{F}_{ex}(\omega)+\hat{f}_N(\omega,\theta)].
\end{align}
The full expression of $\hat{Z}_{out}(\omega,\theta)$ with the gain factor $\mathcal{G}(\omega,\theta)$, the external force signal $\hat{F}_{ex}(\omega)$, and the added force noise $\hat{f}_{N}(\omega,\theta)$ is given by Eq. (\ref{25}) and Eq. (\ref{26}) in  Appendix \ref{app:level5}. The added force noise should be minimized to enhance the force sensitivity. We define the symmetrized added force noise power spectral density \cite{RevModPhys.82.1155, PhysRevA.95.023844}
\begin{align}\label{11}
 \bar{S}_{FF}(\omega,\theta)=&\int\limits_{-\infty}^{\infty}d\omega^\prime\frac{\langle \hat{f}_N^\dag(\omega,\theta)\hat{f}_N(\omega^\prime,\theta)+\hat{f}_N^\dag(\omega^\prime,\theta)\hat{f}_N(\omega,\theta)\rangle}{2}.
\end{align}
The force sensitivity is defined by \cite{PhysRevA.106.023107, PhysRevApplied.17.034020}
\begin{align}\label{11a}
 \mathcal{S}(\omega,\theta)=\sqrt{\bar{S}_{FF}(\omega,\theta)}.
\end{align}
The sensitivity $\mathcal{S}(\omega,\theta)$ corresponds to the value of force signal $F_{ex}(\omega)$ for which the signal-to-noise ratio (SNR) becomes one ({\it i.e.,} $\mathrm{SNR}(\omega,\theta)=|F_{ex}(\omega)|/\sqrt{\bar{S}_{FF}(\omega,\theta)}=1$). Here, $\mathrm{SNR}(\omega,\theta)$ is defined as the ratio of the force signal to the variance of all noise introduced by the system. Consequently, a weaker force signal can be faithfully detected by minimizing the added force noise $\bar{S}_{FF}(\omega,\theta)$. 

The analytical expression of $\bar{S}_{FF}(\omega,\theta)$ can reveal the physical effects of $\theta$ and the squeezing (ICS/IES) parameters on the force sensitivity. We restrict our discussion to $\Delta=0$ for simplicity and to ensure stability of the system as described in Appendix \ref{app:level7}. The derivation of $\bar{S}_{FF}(\omega,\theta)$ for the ICS of the cavity mode is given in Appendix \ref{app:level6a}. The sum total of the shot noise in Eq. (\ref{29ii}) and the backaction noise in Eq. (\ref{29iii}) can be suppressed by optimizing $\bar{S}_{FF}(\omega,\theta)$ w.r.t. $|G|^2$ \cite{PhysRevA.89.053836} that results in
\begin{widetext}
\begin{align}\label{12}
|G_{opt}(\omega,\theta)|^2=&\frac{|\chi_m(\omega)\chi_m^*(-\omega)|^{-1}}{2\omega_m}\sqrt{\frac{A^2+B^2+2AB\cos(2\theta+\phi_d)}{C^2+D^2-2CD\cos(\theta-\psi)}}.
\end{align}
Substituting Eq. (\ref{12}) in Eq. (\ref{29}), we obtain
\begin{align}\label{13}
\bar{S}_{FF}^{opt}(\omega,\theta)=\hbar m\omega_m\gamma &\left[\frac{2|\chi_m(\omega)\chi_m^*(-\omega)|^{-1}\sqrt{[A^2+B^2+2AB\cos(2\theta+\phi_d)][C^2+D^2-2CD\cos(\theta-\psi)]}}{\kappa\gamma\omega_m[\{\kappa\sin(\theta-\psi)-4\Lambda\sin(\theta+\psi+\phi_d)\}^2+\{2\omega\sin(\theta-\psi)\}^2]}\right.\\
&\left.+\frac{2\textrm{Re}[\chi_m^{-1}(\omega)\chi_m^{*^{-1}}(-\omega)][C\{A\cos(\theta+\psi+\phi_d)+B\cos(\theta-\psi)\}-D\{A\cos(2\theta+\phi_d)+B\}]}{\kappa\gamma\omega_m[\{\kappa\sin(\theta-\psi)-4\Lambda\sin(\theta+\psi+\phi_d)\}^2+\{2\omega\sin(\theta-\psi)\}^2]}\right]\nonumber.
\end{align}
\end{widetext}
The exact expressions of $A$, $B$, $C$, and $D$ as functions of different parameters such as $\omega$, $\theta$, $\Lambda$, and $\phi_d$ are given in Appendix \ref{app:level6a}. Here, $\bar{S}_{FF}(\omega,\theta)$ $\mathrm{(\ref{29})}$, $|G_{opt}(\omega,\theta)|^2$ $\mathrm{(\ref{12})}$, and $\bar{S}_{FF}^{opt}(\omega,\theta)$ $\mathrm{(\ref{13})}$ are also the intrinsic functions of $\omega$, $\theta$, $\Lambda$, and $\phi_d$. 
We can neglect the thermal noise contributions from mechanical phonons in Eq. (\ref{29i}), assuming a negligible cryogenic temperature ($\sim10$ mK) environment. The second term in Eq. (\ref{13}) is the simplified shot noise-backaction noise correlation \(\bar{S}_{FF}^{cor}(\omega,\theta)\) given by Eq. (\ref{29iv}), which vanishes for amplitude ($\theta=0$)  or phase ($\theta=0.5\pi$) quadrature measurement of the cavity's output field in the absence of the two-photon drive. 
\subsection{\label{sec:level3a}No squeezing}
In the absence of the degenerate parametric amplification (i.e., $\Lambda=\phi_d=0$), conventional phase quadrature ($\theta=0.5\pi$) readout of the cavity's output field is limited by the SQL for cavity optomechanical weak force sensing. In that case, we can obtain the optimal probe power requirement for SQL-limited weak force sensing
\begin{align}\label{14a}
|G^{SQL}_{opt}(\omega)|^2=\frac{(\omega^2+\kappa^2/4)|\chi_m(\omega)\chi_m^*(-\omega)|^{-1}}{4\kappa\omega_m}
\end{align} 
from Eq. (\ref{12}). A cavity optomechanical system in the so-called bad-cavity or unresolved sideband regime ($\kappa\gg\omega_m$) is more sensitive, as there is less time delay in the cavity's response to weak external forces. Such systems are ideally suited for weak force sensing in cavity optomechanics \cite{PhysRevA.89.053836, Motazedifard_2016}. Substituting $\theta=0.5\pi$ and $\Lambda=\phi_d=\psi=0$ in Eq. (\ref{13}), we obtain 
\begin{align}\label{14b}
\bar{S}_{FF}^{SQL}(\omega)=&\hbar m\omega_m\gamma[|\chi_m(\omega)\chi_m^*(-\omega)|^{-1}/(\gamma\omega_m)],
\end{align}
which is the expression of the SQL for $\bar{S}_{FF}(\omega)$ and represents the lowest value of $\bar{S}_{FF}^{opt}(\omega,0.5\pi)$ for arbitrary values of $\omega$. However, measuring a generalized homodyne quadrature, as given by Eq. (\ref{9}), introduces quantum correlations between the amplitude and phase quadratures of the cavity's output field. Quantum correlation between the quadratures can further correlate the cavity input shot noise to the radiation-pressure backaction noise, thereby effectively reducing the total added force noise. An appropriate homodyne angle $\theta$ can reduce the added force noise below SQL within a certain frequency band similar to the ponderomotive squeezing \cite{PhysRevX.7.021008, PhysRevA.110.043512}.

The advantage of quantum correlations for enhanced weak force sensing is illustrated in Fig. \ref{fig:fig2}. In Fig. \ref{fig:fig2} (a), we show that the $\bar{S}_{FF}^{opt}(\omega)$ lower than that at SQL (dotted black line) can be achieved by the cavity output quadrature readout with a homodyne angle $\theta=0.9\pi$ ($\theta=0.1\pi$) for $\omega<\omega_m$ ($\omega>\omega_m$) as shown by the solid red (dashed blue) line. Off-resonant ($\omega\neq\omega_m$) sub-SQL sensitivity over a broad frequency range is beneficial for broadband cavity optomechanical weak force sensing. Moreover, we illustrate the $\bar{S}_{FF}^{opt}(\omega)$ as a function of the normalized frequency $\omega/\omega_m$ and homodyne angle $\theta$ in Fig. \ref{fig:fig2} (b) to provide a lucid visualization of the dependence of sensitivity on these parameters. We observe that the weak force sensitivity gradually increases in the lower- ($\omega<\omega_m$) and the higher- ($\omega>\omega_m$) frequency regions by tuning the homodyne angle away from $\theta=0.5\pi$ towards $\theta=\pi$ and $\theta=0$, respectively. Since the correlation noise $\bar{S}_{FF}^{cor}(\omega,\theta)$ and the gain factor $\mathcal{G}(\omega,\theta)$ in Eq. (\ref{10}) vanish at $\theta=0,\pi$ for $\Lambda=\phi_d=0$, we use $\theta=0.9\pi$ and $\theta=0.1\pi$ in Fig. \ref{fig:fig2} (a). One can minimize $\bar{S}_{FF}^{opt}(\omega)$ by optimizing $\theta$. In the absence of a parametric drive 
\begin{align}\label{15}
\theta_{opt}(\omega)=&\cos^{-1}\left[\frac{-Re\{\chi_m^{-1}(\omega)\chi_m^{*^{-1}}(-\omega)\}}{|\chi_m(\omega)\chi_m^*(-\omega)|^{-1}}\right],
\end{align}
which provides us $\bar{S}_{FF}^{opt}(\omega)|_{\theta=\theta_{opt}}=\hbar m\omega_m\gamma(\omega/\omega_m)$ that is shown with a dash-dotted magenta line and the inset in Fig. \ref{fig:fig2} (a). Thus, optimal weak force sensitivity lower than $\bar{S}_{FF}^{SQL}(\omega_m)\approx\hbar m\omega_m\gamma$ at the mechanical resonant frequency can be achieved only in the lower-frequency region, $\omega<\omega_m$. Fig. \ref{fig:fig2} (c) elucidates the variation of $\theta_{opt}(\omega)$ for different values of normalized frequency $\omega/\omega_m$. We observe that $\theta_{opt}(\omega)$ is very close to $\pi$ (0) for $\omega<\omega_m$ ($\omega>\omega_m$), which agrees well with the result shown in Fig. \ref{fig:fig2} (b).
\subsection{\label{sec:level3b}Intra-cavity squeezing}
\begin{figure}[b!]
    \centering
    \includegraphics[width=0.45\textwidth]{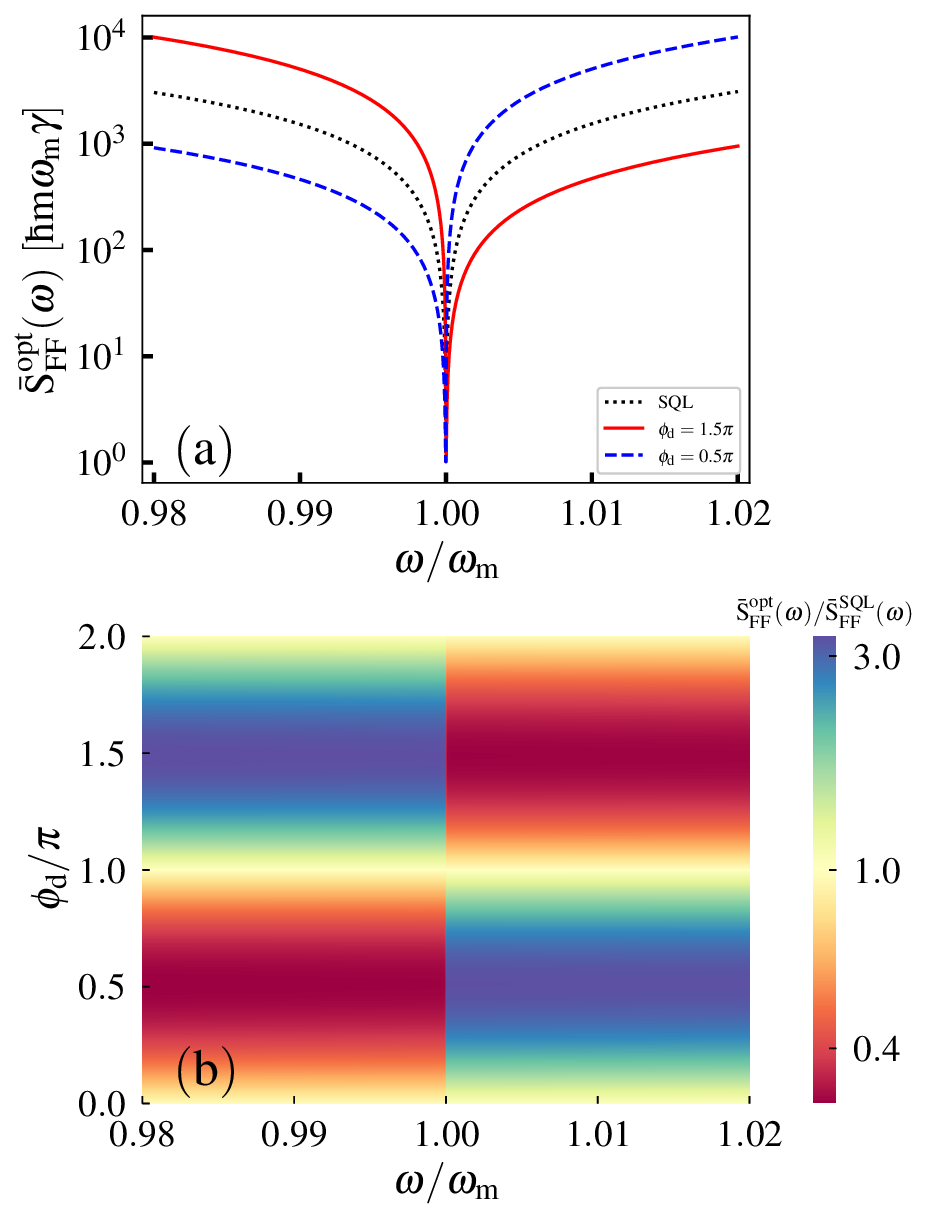}
    \caption{(Color online) (a) The optimal added force noise power spectral density $\bar{S}_{FF}^{opt}(\omega)$ (in the units of $\hbar m\omega_m\gamma$) vs $\omega/\omega_m$ for SQL, $\theta=0.5\pi$, $\Lambda=0.1\kappa$, and different values of $\phi_d$. (b) The surface plot shows the ratio of the optimal added force noise power spectral density to that at SQL as a function of $\phi_d$ and $\omega/\omega_m$ for $\theta=0.5\pi$ and $\Lambda=0.1\kappa$. The other parameters used for this figure are the same as those used in Fig. \ref{fig:fig2}.}
    \label{fig:fig3}
\end{figure}
This section discusses ICS-enhanced weak force detection while using the variational homodyne detection technique in the absence of IES (\(r_e=\phi_e=0\)). We utilize the optimal linearized optomechanical coupling \( |G_{opt}(\omega,\theta)|^2 \) and the corresponding optimal added force noise power spectral density \( \bar{S}_{FF}^{opt}(\omega,\theta) \), as presented in Eqs. (\ref{12}) and (\ref{13}), respectively. We first consider the homodyne phase quadrature readout (\(\theta = 0.5\pi\)) of the cavity's output field, assuming no parametric drive is applied. Under these conditions, the amplitude and phase quadrature noise remain uncorrelated, resulting in conventional SQL as depicted by the dotted black line in Fig. \ref{fig:fig3}(a). We then introduce a parametric drive with strength \( \Lambda = 0.1\kappa \) to ensure system stability (see Appendix \ref{app:level7}). By tuning the phase of the parametric drive \( \phi_d \), we can manipulate the quantum correlation noise \( \bar{S}_{FF}^{cor}(\omega) \). The solid red line shows that the optimal added force noise falls below the SQL for frequencies greater than the mechanical oscillator frequency $\omega_m$ when \( \phi_d=1.5\pi \). On the contrary, the dashed blue line shows that $\phi_d=0.5\pi$ helps to beat SQL for frequencies less than $\omega_m$. A recent study has also demonstrated similar results in a parametrically driven cavity optomechanical system \cite {Zhao2019}.

\begin{figure}[t!]
    \centering
    \includegraphics[width=0.48\textwidth]{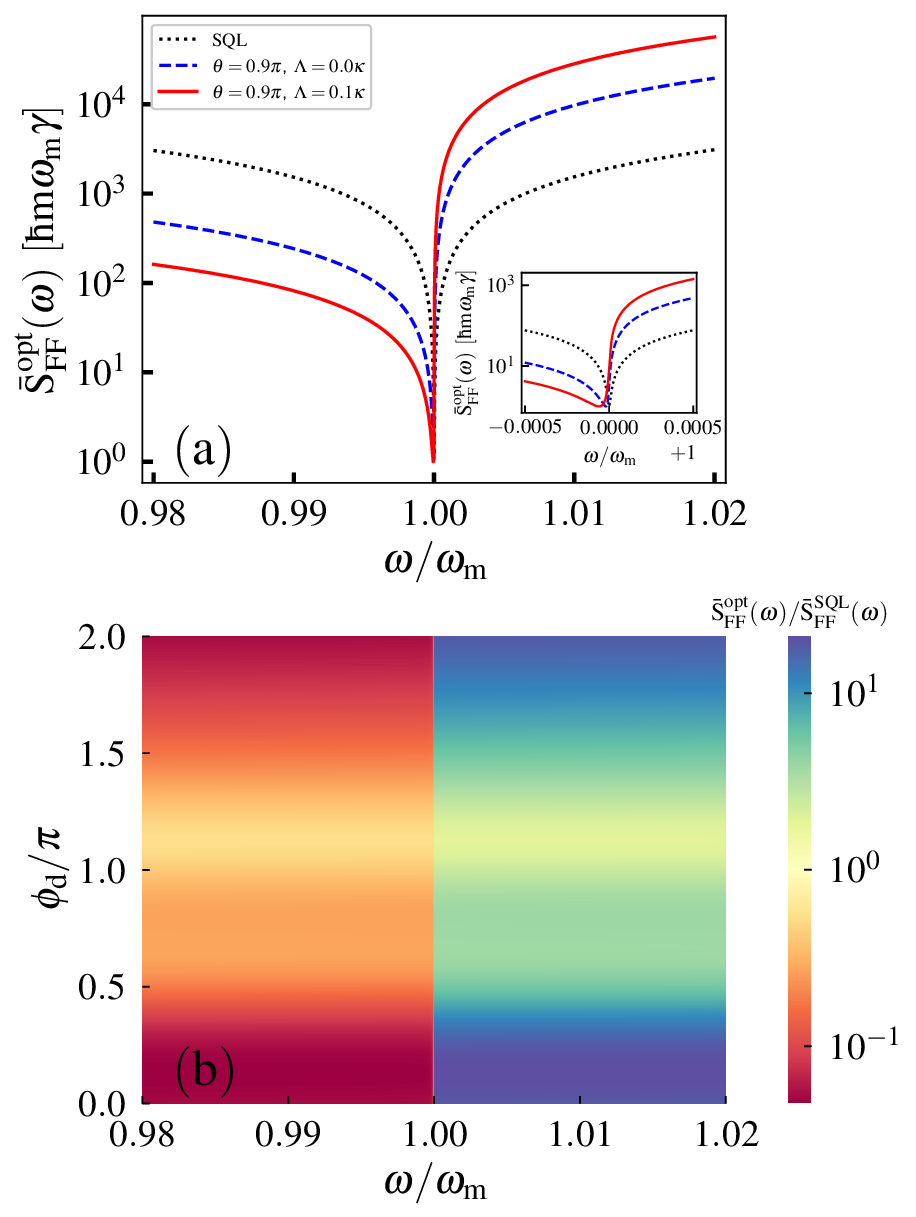}
    \caption{(Color online) (a) The optimal added force noise power spectral density $\bar{S}_{FF}^{opt}(\omega)$ (in the units of $\hbar m\omega_m\gamma$) vs $\omega/\omega_m$ for SQL, $\theta=0.9\pi$, $\phi_d=0$, and different values of $\Lambda$. The inset shows the magnified view in the vicinity of $\omega=\omega_m$. (b) The surface plot shows the ratio of the optimal added force noise power spectral density to that at SQL as a function of $\phi_d$ and $\omega/\omega_m$ for $\theta=0.9\pi$ and $\Lambda=0.1\kappa$. The other parameters used for this figure are the same as those used in Fig. \ref{fig:fig2}.}
    \label{fig:fig4}
\end{figure} 
These observations are supported by a surface plot in Fig. \ref{fig:fig3} (b), which shows the dependence of $\bar{S}_{FF}^{opt}(\omega)$ on the parametric drive phase $\phi_d$ and $\omega/\omega_m$. From this plot, it is evident that $\bar{S}_{FF}^{opt}(\omega)$ can drop below the SQL when $\omega<\omega_m$ and $0 <\phi_d<\pi$ due to the quantum correlation noise \( \bar{S}_{FF}^{cor}(\omega) \). Conversely, for $\omega>\omega_m$, $\bar{S}_{FF}^{opt}(\omega)$ can be reduced below the SQL when $\pi<\phi_d<2\pi$.  Furthermore, \(S_{FF}^{opt}(\omega)\) is minimum around \(\phi_d=0.5\pi\) and $\phi_d=1.5\pi$ when $\omega<\omega_m$ and $\omega>\omega_m$, respectively. However, the system can become unstable for $\Lambda=0.1\kappa$, $0<\phi_d<\pi$, and higher values of \( |G_{opt}(\omega,0.5\pi)|\) as shown in Fig. \ref{fig:fig13} (b) and (d) of Appendix \ref{app:level7}.
\begin{figure}[t!]
    \centering
    \includegraphics[width=0.4\textwidth]{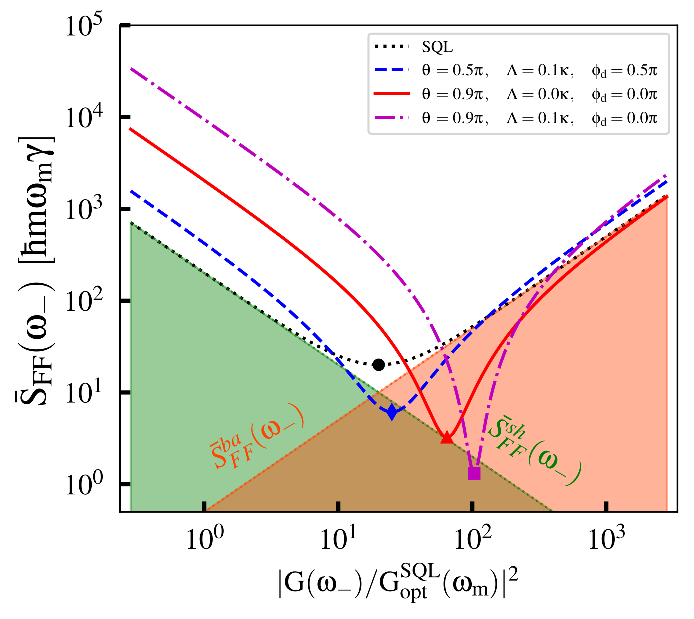}
    \caption{(Color online) The added force noise power spectral density $\bar{S}_{FF}(\omega_-)$ (in the units of $\hbar m\omega_m\gamma$) vs the normalized probe power $|G(\omega_-)/G_{opt}^{SQL}(\omega_m)|^2$ for a frequency $\omega_-=\omega_m-10\gamma$ with different values of $\theta$, $\Lambda$, and $\phi_d$. The other parameters used for this plot are the same as those used in Fig. \ref{fig:fig2}.}
    \label{fig:fig5}
\end{figure}

We now investigate whether the combined effects of quantum correlations and ICS-induced quantum squeezing can further enhance the force sensitivity. As demonstrated in Sec.\ref{sec:level3a}, choosing $\theta=0.9\pi$ ($\theta=0.1\pi$) enhances the force sensitivity for $\omega<\omega_m$ ($\omega>\omega_m$). Interestingly, we observe in Fig. \ref{fig:fig4} (a) that a lower sub-SQL $\bar{S}_{FF}^{opt}(\omega)$ can be achieved with $\Lambda=0.1\kappa$ (solid red line) than that achievable with $\Lambda=0$ (dashed blue line) for $\theta=0.9\pi$, \(\phi_d=0\), and $\omega<\omega_m$. Thus, the prametric drive strength \(\Lambda\) manipulates the quantum correlation noise \( \bar{S}_{FF}^{cor}(\omega) \)  for \(\theta=0.9\pi\), \(\phi_d=0\), and \(\omega<\omega_m\). The inset in Fig. \ref{fig:fig4} (a) shows the behaviour of $\bar{S}_{FF}^{opt}(\omega)$ in the vicinity of the mechanical resonant frequency $\omega_m$, which does not differ much from the main figure. The surface plot in Fig. \ref{fig:fig4} (b) depicts that $\phi_d=2n\pi$ ($n=0,1,2,..$) optimally enhances the force sensitivity when $\theta=0.9\pi$, $\Lambda=0.1\kappa$, and \(\omega<\omega_m\). Increasing $\Lambda>0.1\kappa$ results in more arbitrary behavior in the dependence of $\bar{S}_{FF}^{opt}(\omega)$ on $\phi_d$ and $\omega/\omega_m$ as shown in Fig. \ref{fig:fig9} of Appendix \ref{app:level6a}. Moreover, a larger value of $\Lambda$ and \(0<\phi_d<\pi\) can destabilize the system at higher \(|G_{opt}(\omega,\theta)|\) values (see Appendix \ref{app:level7}). The dependence of $\bar{S}_{FF}^{opt}(\omega)$ on $\phi_d$ and $\omega/\omega_m$ for $\Lambda=0.1\kappa$, $\theta=0.1\pi$, and $\omega>\omega_m$ are shown in Fig. \ref{fig:fig10} (a) and (b) of Appendix \ref{app:level6a}, which depict the patterns opposite to Fig. \ref{fig:fig4} (a) and (b) around $\omega=\omega_m$. 

Finally, we discuss the physical mechanisms underlying the remarkable enhancement of weak force sensitivity via quantum correlations and the ICS. In Fig. \ref{fig:fig5}, we show the probe power requirements for optimal force sensing at a frequency $\omega_-=\omega_m-10\gamma$ with various values of $\theta,\Lambda,$ and $\phi_d$ discussed earlier in this Sec. \ref{sec:level3b}. The probe power $|\Omega|^2$ is directly proportional to $|G(\omega)|^2$ as evident from Eq. (\ref{5a}). The thick-dotted black line delineates the SQL condition (\textit{i.e.,} $\theta=0.5\pi$ and $\Lambda=\phi_d=0$). The thin-dotted green (thin-dotted orange) line shows the contribution of shot noise (backaction noise) in $\bar{S}_{FF}^{SQL}(\omega_-)$. The parameters used for the other curves are listed in the figure legend. We observe that the force sensitivity at an off-resonant frequency $\omega_-$ can be optimally enhanced via quantum correlation and ICS by increasing the probe power \cite{PhysRevX.7.021008}.  In Fig. \ref{fig:fig5}, we show that $|G_{opt}(\omega_-)|^2$ values are 20 (circle), 25 (diamond), 65 (triangle), and 103 (square) times that of $|G_{opt}^{SQL}(\omega_m)|^2$ for the dotted black, dashed blue, solid red, and dash-dotted magenta lines for optimal sub-SQL force detection, respectively. However, the blue diamond corresponds to a $|G(\omega_-)/G_{opt}^{SQL}(\omega_m)|$ value within the unstable region for $\Lambda/\kappa=0.1$ and $\phi_d=0.5\pi$ in Fig. \ref{fig:fig13} (d) of Appendix \ref{app:level7}, which is experimentally unfeasible. The relative quantum advantage for optimal sub-SQL force sensitivity w.r.t. the SQL can be defined in the decibel units as 
\begin{align}\label{16}
\delta S(\omega,\theta)=&10\times log_{10}\left(\frac{\bar{S}_{FF}^{opt}(\omega,\theta)}{\bar{S}_{FF}^{SQL}(\omega)}\right).
\end{align}
The quantum advantages for the dashed blue, solid red, and dash-dotted magenta lines at $\omega_-$ are $-11.85$ dB, $-18.19$ dB, and $-27.26$ dB, respectively. The solid red line shows that choosing $\theta=0.9\pi$ increases shot noise, while reducing the added force noise by beating the backaction noise beyond SQL via quantum correlation. The dash-dotted magenta line depicts a substantial reduction in backaction noise and a marginal reduction in shot noise below SQL via ICS, with an increased $|G(\omega_-)|^2$ value. Here, both quantum correlation and ICS are leveraged to reduce $\bar{S}_{FF}^{opt}(\omega_-)$ beyond SQL. The probe power requirements for $\Lambda=0.1\kappa$, $\theta=0.1\pi$, and $\omega_+=\omega_m+10\gamma$, are shown in Fig. \ref{fig:fig10} (c) of Appendix \ref{app:level6a}.
\subsection{\label{sec:level3c}Injected external squeezing}
In this section, we consider a broadband squeezed vacuum injected into the cavity with a squeezing parameter $r_e$ and a squeezing angle $\phi_e$ \cite{Murch2013} for IES-enhanced weak force sensing in the absence of ICS ($\Lambda=\phi_d=0$). Restricting our discussion to $\Delta = 0$, optimization of $\bar{S}_{FF}(\omega)$ given by Eq. (\ref{31}) in Appendix \ref{app:level6b} w.r.t. $|G|^2$ gives us
\begin{widetext}
\begin{align}\label{17}
|G_{opt}(\omega,\theta)|^2=&\frac{(\omega^2+\kappa^2/4)|\chi_m(\omega)\chi_m^*(-\omega)|^{-1}}{2\omega_m[2\kappa\sin(\theta)]}\sqrt{\frac{[\cos(2\theta-\phi_e)\sinh(2r_e)+\cosh(2r_e)]}{\cos(\phi_e)\sinh(2r_e)+\cosh(2r_e)}}. 
\end{align}
Substituting Eq. (\ref{17}) in Eq. (\ref{31}), we obtain
\begin{align}\label{18}
\bar{S}_{FF}^{opt}(\omega,\theta)=\hbar m\omega_m\gamma &\left[\frac{|\chi_m(\omega)\chi_m^*(-\omega)|^{-1}\sqrt{[\cos(2\theta-\phi_e)\sinh(2r_e)+\cosh(2r_e)][\cos(\phi_e)\sinh(2r_e)+\cosh(2r_e)]}}{\gamma\omega_m\sin(\theta)}\right.\nonumber\\
&\left.+\frac{Re[\chi_m^{-1}(\omega)\chi_m^{*^{-1}}(-\omega)][\cos
(\theta-\phi_e)\sinh(2r_e)+\cos(\theta)\cosh(2r_e)]}{\gamma\omega_m\sin(\theta)}\right].
\end{align}
\end{widetext}

\begin{figure}[b!]
    \centering
    \includegraphics[width=0.45\textwidth]{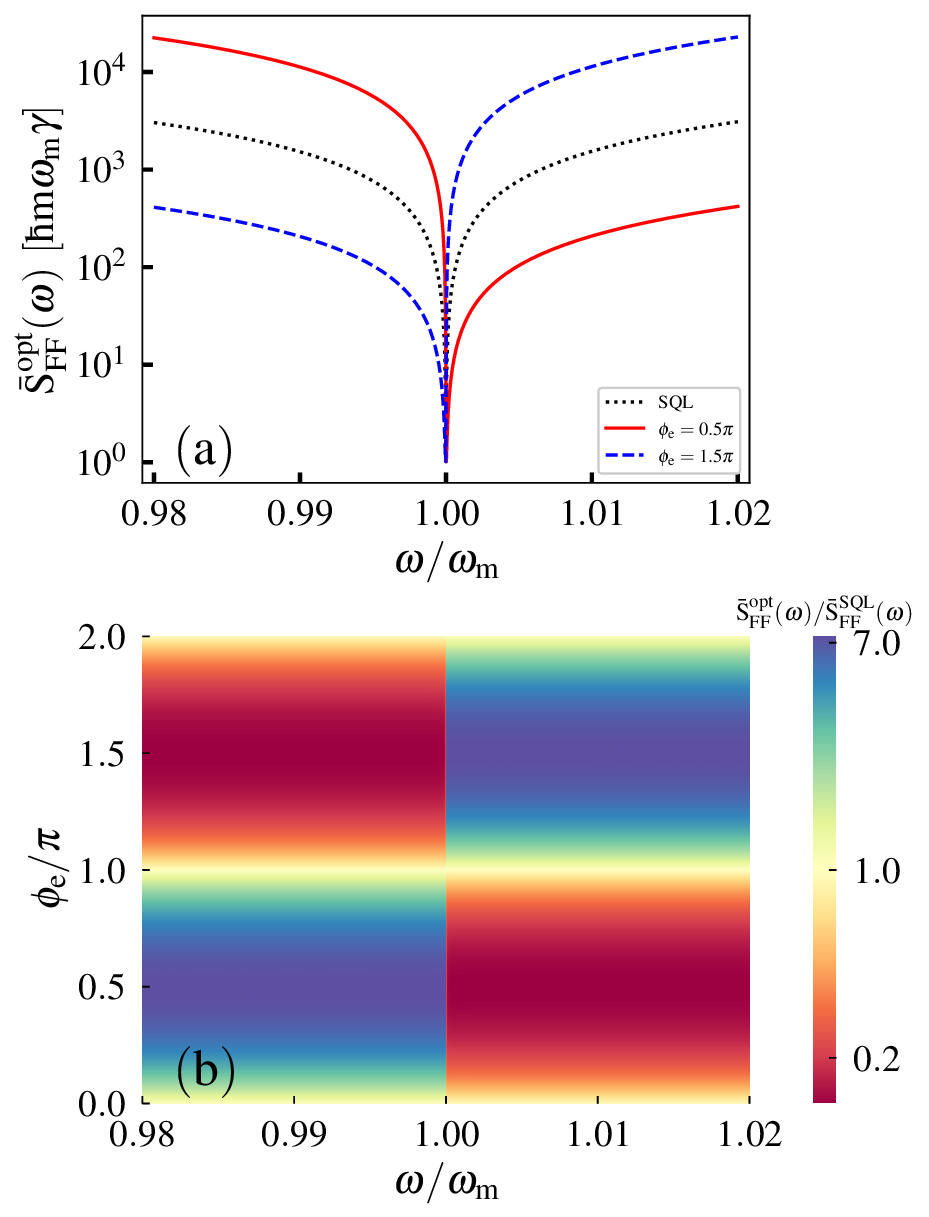}
    \caption{(Color online) (a) The optimal added force noise power spectral density $\bar{S}_{FF}^{opt}(\omega)$ (in the units of $\hbar m\omega_m\gamma$) vs $\omega/\omega_m$ for SQL, $\theta=0.5\pi$, $r_e=1$, and different values of $\phi_e$. (b) The surface plot shows the ratio of the optimal added force noise power spectral density to that at SQL as a function of $\phi_e$ and $\omega/\omega_m$ for $\theta=0.5\pi$ and $r_e=1$. The other parameters used for this figure are the same as those used in Fig. \ref{fig:fig2}.}
    \label{fig:fig6}
\end{figure}

\noindent Here, $\bar{S}_{FF}(\omega,\theta)$ $\mathrm{(\ref{31})}$, $|G_{opt}(\omega,\theta)|^2$ $\mathrm{(\ref{17})}$, and $\bar{S}_{FF}^{opt}(\omega,\theta)$ $\mathrm{(\ref{18})}$ are the intrinsic functions of $\omega$, $\theta$, $r_e$, and $\phi_e$. In Fig. \ref{fig:fig6} (a), the dashed blue line corresponds to $\phi_e=0.5\pi$, which enables sub-SQL weak force sensing at frequencies below $\omega_m$ for \(\theta=0.5\pi\). Likewise, the solid red line shows that $\phi_e=1.5\pi$ enhances the weak force sensitivity beyond SQL for  \(\theta=0.5\pi\) and $\omega>\omega_m$. The surface plot in Fig. \ref{fig:fig6} (b) illustrates the force sensitivity as a function of $\phi_e$ and $\omega/\omega_m$ for $\theta=0.5\pi$ and $r_e=1$. This figure suggests that lower sub-SQL force sensitivity (deep-red region) can be achieved at any frequency above (below) the mechanical resonant frequency $\omega_m$ due to quantum correlation introduced by the squeezing angle $\phi_e=0.5\pi$ ($\phi_e=1.5\pi$). These observations can be attributed to the fact that tuning $\phi_e$ with a fixed $r_e=1$ and $\theta=0.5\pi$ correlates the amplitude and phase quadratures of the cavity's output field. This correlation results in nonzero $\bar{S}_{FF}^{cor}(\omega)$ that can reduce the total $\bar{S}_{FF}^{opt}(\omega)$ beyond SQL.
\begin{figure}[t!]
    \centering
    \includegraphics[width=0.45\textwidth]{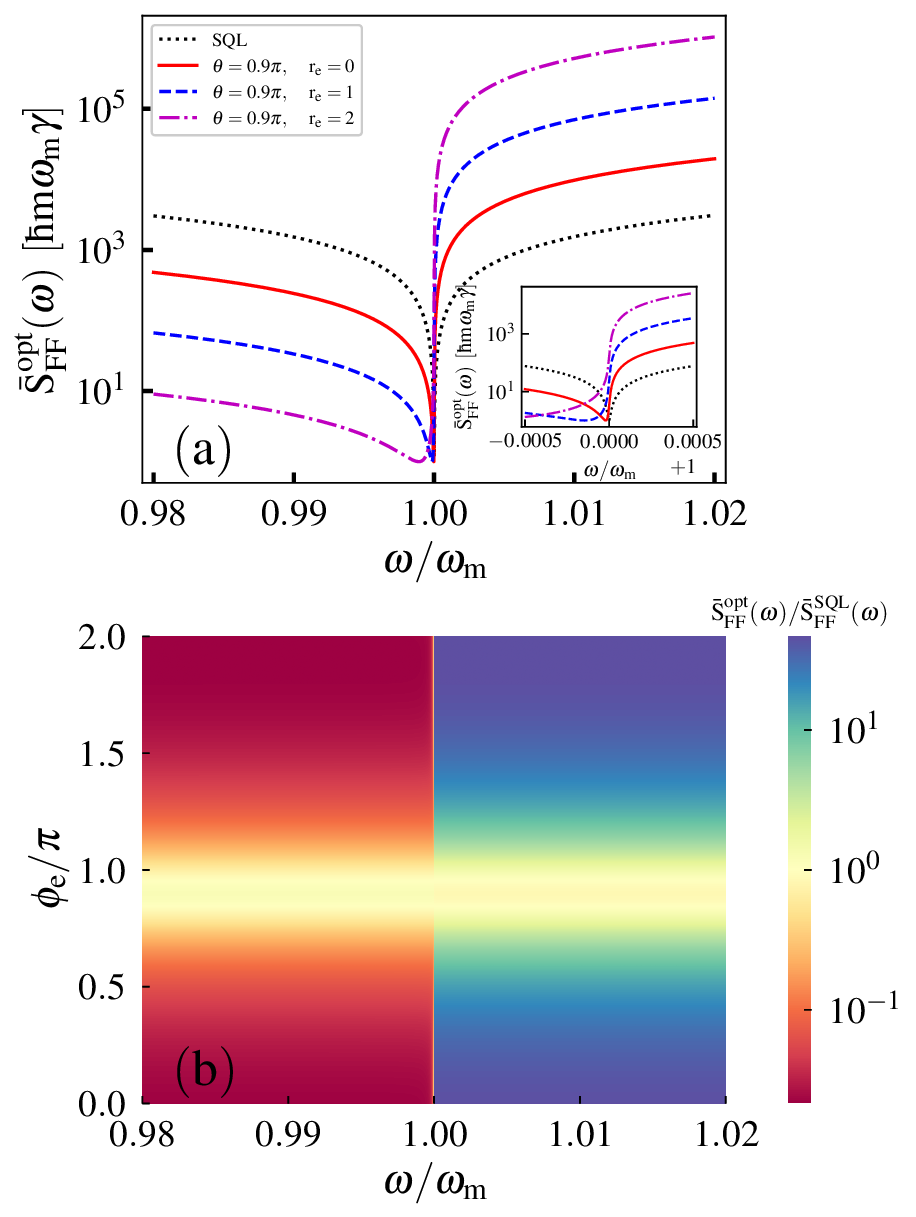}
    \caption{(Color online) (a) The optimal added force noise power spectral density $\bar{S}_{FF}^{opt}(\omega)$ (in the units of $\hbar m\omega_m\gamma$) vs $\omega/\omega_m$ for SQL, $\theta=0.9\pi$, $\phi_e=0$, and different values of $r_e$. The inset shows the magnified view in the vicinity of $\omega=\omega_m$. (b) The surface plot shows the ratio of the optimal added force noise power spectral density to that at SQL as a function of $\phi_e$ and $\omega/\omega_m$ for $\theta=0.9\pi$ and $r_e=1$. The other parameters used for this figure are the same as those used in Fig. \ref{fig:fig2}.}
    \label{fig:fig7}
\end{figure}

Moving on, we discuss the sub-SQL weak force sensing for $\omega<\omega_m$ via the homodyne quadrature readout with $\theta=0.9\pi$ and different values of IES parameters. In Fig. \ref{fig:fig7} (a), we show that $\bar{S}_{FF}^{opt}(\omega)$ keeps decreasing below SQL for increasing values of $r_e$ with $\theta=0.9\pi$ and $\phi_e=2n\pi$ ($n=0,1,2,..$). However, this is not true for frequencies very close to the resonant mechanical frequency $\omega_m$ as shown in the inset of Fig. \ref{fig:fig7} (a). In Fig. \ref{fig:fig7} (b), the surface plot shows that the force sensitivity can be optimized with $\phi_e=2n\pi$, \((n=0,1,2,..)\) for $r_e=1$ when $\theta=0.9\pi$ and \(\omega<\omega_m\). Thus, tuning \( r_e\) can control \( \bar{S}_{FF}^{cor}(\omega)\) to achieve sub-SQL \( \bar{S}_{FF}^{opt}(\omega) \) values when \(\theta=0.9\pi\), \(\phi_e=0\) \((2n\pi)\), and \(\omega<\omega_m\). We observe similar behaviour of $\bar{S}_{FF}^{opt}(\omega)$ for $\theta=0.1\pi$, $\phi_e=2n\pi$, and \(\omega>\omega_m\) as shown in Fig. \ref{fig:fig11} (a) and (b) of Appendix \ref{app:level6b}.
\begin{figure}[t!]
    \centering
    \includegraphics[width=0.4\textwidth]{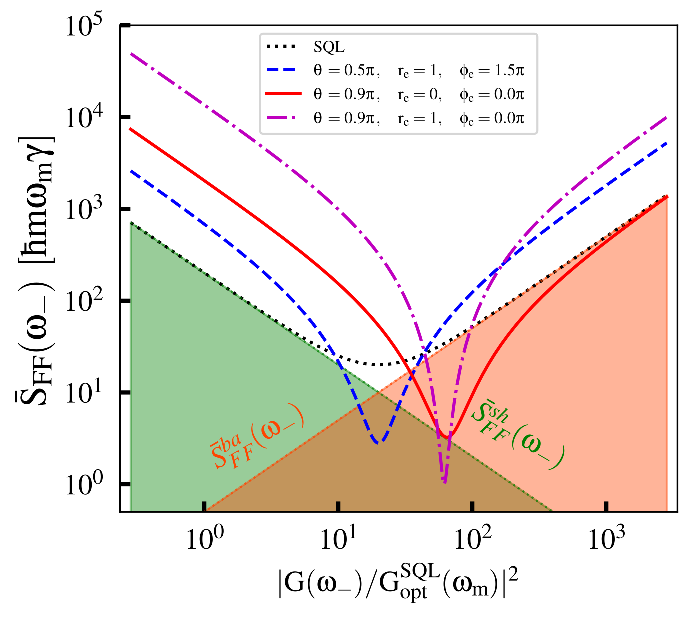}
    \caption{(Color online) The added force noise power spectral density $\bar{S}_{FF}(\omega_-)$ (in the units of $\hbar m\omega_m\gamma$) vs the normalized probe power $|G(\omega_-)/G_{opt}^{SQL}(\omega_m)|^2$ for a frequency $\omega_-=\omega_m-10\gamma$ with different values of $\theta$, $r_e$, and $\phi_e$. The other parameters used for this figure are the same as those used in Fig. \ref{fig:fig2}.}
    \label{fig:fig8}
\end{figure}

The probe powers required for optimally enhanced weak force sensing with various homodyne angles and IES parameters at a frequency \(\omega_-=\omega_m-10\gamma\) are illustrated in Fig. \ref{fig:fig8}. The probe powers shown in the figure are normalized with respect to $|G_{opt}^{SQL}(\omega_m)|^2$. This figure provides a deeper understanding of the physical mechanism underlying the noise reduction. We show that the force sensitivity can be enhanced by only controlling the IES parameters $r_e$ and $\phi_e$ without increasing the cavity probe power significantly for a particular homodyne angle $\theta$. For the homodyne phase quadrature readout ($\theta=0.5\pi$), increasing the value of $r_e$ reduces $\bar{S}_{FF}^{opt}(\omega)$ by keeping $\phi_e=1.5\pi$ and the cavity probe power fixed. However, a higher probe power is required for optimizing the force sensitivity for the homodyne quadrature readout with $\theta=0.9\pi$. The force sensitivity can be further enhanced by only increasing $r_e$ with fixed $\phi_e=0$ \((2n\pi)\) and $\theta=0.9\pi$ for $\omega<\omega_m$, while slightly reducing the probe power. The thick-dotted black line represents the SQL parameter condition ($\theta=0.5\pi$, $r_e=0$, and $\phi_e=0$). The thin-dotted green and the thin-dotted orange lines show the contributions of the shot noise and the backaction noise in $\bar{S}_{FF}^{SQL}(\omega_-)$, respectively. The parameters for the other curves are mentioned in the figure legend. The exact optimal probe power required for the SQL and dashed blue line is 20 times that of $|G_{opt}^{SQL}(\omega_m)|^2$. Similarly, one would require probe powers of the order of 65 times and 62 times that of $|G_{opt}^{SQL}(\omega_m)|^2$ for the solid red and dash-dotted magenta lines, respectively. The quantum advantage for weak force sensing for the dashed blue, solid red, and dash-dotted magenta lines w.r.t. SQL are $-19.67$ dB, $-18.19$ dB, and $-29.88$ dB, respectively. The solid red line in Fig. \ref{fig:fig8} suggests that the quantum correlation introduced by the homodyne angle $\theta=0.9\pi$ enables sub-SQL force sensing by reducing backaction noise below the SQL in the absence of IES ($r_e=\phi_e=0$). In contrast to the ICS, the IES substantially reduces shot noise below SQL with lower probe powers, as shown by the dashed blue and dash-dotted magenta lines. The dash-dotted magenta line demonstrates that quantum correlation can be applied in conjunction with IES to achieve a remarkable enhancement in force sensitivity with a relatively lower probe power than the ICS case. Fig. \ref{fig:fig11} (c) shows very similar probe power requirements and quantum advantages for $\omega_+=\omega_m+10\gamma$, $\theta=0.1\pi$, and $r_e=1$.
\section{\label{sec:level4}Conclusion}
In this paper, we have investigated the enhancement of the weak force sensitivity of an electromechanical system through variational homodyne readout of the cavity's output quadrature and quantum squeezing of the cavity mode. The introduction of quantum correlations via homodyne quadrature readouts with $0.5\pi<\theta<\pi$ ($0<\theta<0.5\pi$) enhances force sensitivity for $\omega>\omega_m$ ($\omega<\omega_m$) while increasing the shot noise and reducing the backaction noise beyond SQL. For instance, we have shown that the cavity's output quadrature measurement with homodyne angles $\theta=0.9\pi$ and $\theta=0.1\pi$ enables sub-SQL force sensing in our proposed electromechanical system for $\omega<\omega_m$ and $\omega>\omega_m$, respectively. A frequency-dependent variational homodyne angle $\theta_{opt}(\omega)$ can enable optimal sub-SQL weak force sensing in the electromechanical system. Moreover, our results substantiate that quantum squeezing of the cavity mode via ICS or IES can further enhance force sensitivity by reducing noise. Squeezing the cavity mode via ICS or IES reduces both the shot noise and the backaction noise beyond the SQL. However, a larger two-photon/parametric drive for ICS can destabilize the system. We have also observed that IES substantially reduces shot noise at lower probe powers than ICS for sub-SQL force sensing at an off-resonant frequency $\omega_{\pm}=\omega_m\pm10\gamma$. Therefore, IES can be considered preferable to ICS for squeezing enhanced variational homodyne weak force sensing. 

This paper demonstrates that variational homodyne quadrature readout and quantum squeezing together can significantly enhance the weak force sensitivity of a cavity optomechanical sensor. Our work exploits the advantages of both quantum correlation and quantum squeezing to achieve broadband sub-SQL force sensitivity in cavity optomechanics. Although we have proposed the experimental implementation of our model in an electromechanical system \cite{PhysRevLett.121.243601, Clark2017ER}, any other cavity optomechanical system can be utilized for this purpose \cite{Mason2019, Fogliano2021}. The proposed squeezing enhanced variational homodyne readout schemes for weak force sensing can be integrated with feedback control \cite{PhysRevApplied.17.034020}, Kerr nonlinearity \cite{9w37-r1y8}, quadratic optomechanical coupling \cite{Zhang:24}, and non-hermitian systems \cite{kmtx-7x9d}, and so on. The combination of both ICS and IES is also another interesting direction to explore \cite{PhysRevLett.114.093602, PhysRevLett.133.233605, PhysRevA.110.063520}. 
\section*{\label{sec:level4a}Acknowledgment}
M.M.M. and T.N.D. gratefully acknowledge funding by the Department of Science and Technology, Anusandhan National Research Foundation, Government of India (Grant No. CRG/2023/001318). S.D. thanks the Okinawa Institute of Science and Technology Graduate University for financial support.
\section*{\label{sec:level4b}Data availability}
The data that support the findings of this article are not
publicly available. The data are available from the authors
upon reasonable request.
\appendix
\begin{widetext}
\section{\label{app:level5}The homodyne cavity output field quadrature}
In the Fourier frequency domain applying $\hat{\mathcal{O}}^\dag(\omega)=[\hat{\mathcal{O}}(-\omega)]^\dag$ on Eqs. (\ref{6}) and (\ref{7}) of Sec. \ref{sec:level2}, we obtain
\begin{align}
\frac{\hat{a}^\dag(\omega)}{\chi^*_c(-\omega)} &= -iG[\hat{b}(\omega)+\hat{b}^\dag(\omega)]+2\Lambda\hat{a}(\omega)e^{i\phi_d}+\sqrt{\kappa}\hat{a}^\dag_{in}(\omega),\label{19i}\\
\frac{\hat{b}^\dag(\omega)}{\chi^*_m(-\omega)}&=-i[G\hat{a}^\dag(\omega)+G^*\hat{a}(\omega)]-i\frac{\hat{F}_{ex}(\omega)}{\hbar}x_{ZPF}+\sqrt{\gamma}\hat{b}^\dag_{in}(\omega).\label{19ii}
\end{align}
 Using Eq. (\ref{7}) of Sec. \ref{sec:level2} and Eq. (\ref{19ii}), one can calculate
\begin{align}\label{20}
\hat{b}(\omega)+\hat{b}^\dag(\omega)=&\left[\frac{2\omega_m}{\omega_m^2+(\gamma/2-i\omega)^2}\right][\{G\hat{a}^\dag(\omega)+G^*\hat{a}(\omega)\}+\frac{\hat{F}_{ex}(\omega)}{\hbar}x_{ZPF}]+\sqrt{\gamma}[\chi_m(\omega)\hat{b}_{in}(\omega)+\chi_m^*(-\omega)\hat{b}_{in}^\dag(\omega)].
\end{align}
Substituting Eq. (\ref{20}) in Eq. (\ref{6}) of Sec. \ref{sec:level2} and Eq. (\ref{19i}), we have 
\begin{align}\label{21}
&\begin{bmatrix}
    \chi^{-1}(\omega) & [i\Sigma(\omega)e^{2i\psi}-2\Lambda e^{-i\phi_d}]\\
    -[i\Sigma(\omega)e^{-2i\psi}+2\Lambda e^{i\phi_d}] & {\chi^*}^{-1}(-\omega)
\end{bmatrix}
\begin{bmatrix}
    \hat{a}(\omega)\\
    \hat{a}^\dag(\omega)
\end{bmatrix}
=&
\begin{bmatrix}
    -\frac{i\Sigma(\omega)x_{ZPF}\hat{F}_{ex}(\omega)}{\hbar G^*}+iG\hat{X}_0(\omega)+\sqrt{\kappa}\hat{a}_{in}(\omega)\\
    \frac{i\Sigma(\omega)x_{ZPF}\hat{F}_{ex}(\omega)}{\hbar G}-iG^*\hat{X}_0(\omega)+\sqrt{\kappa}\hat{a}^\dag_{in}(\omega)
\end{bmatrix},
\end{align}
where $\chi(\omega)=[\chi_c^{-1}(\omega)+i\Sigma(\omega)]^{-1}$ is the effective susceptibility of the cavity mode, $\Sigma(\omega)=-2|G|^2\omega_m/[\omega_m^2+(\kappa/2-i\omega)^2]=\Sigma^*(-\omega)$ is the mechanical self-energy, and $\hat{X}_0(\omega)=\sqrt{\gamma}[\chi_m(\omega)\hat{b}_{in}(\omega)+\chi_m^*(-\omega)\hat{b}_{in}^\dag(\omega)]$ is proportional to the steady-state mechanical displacement in absence of the optomechanical interaction. Solving Eq. (\ref{21}), we obtain 
\begin{align}\label{22}
    \hat{a}_{out}(\omega)=&-\frac{[D(\omega)-\kappa{\chi^*}^{-1}(-\omega)]\hat{a}_{in}(\omega)+\kappa[i\Sigma(\omega)e^{2i\psi}-2\Lambda e^{-i\phi_d}]\hat{a}_{in}^\dag(\omega)}{D(\omega)}\nonumber\\
    &-\frac{i\sqrt{\kappa}[{\chi_c^*}^{-1}(-\omega)e^{i\psi}-2\Lambda e^{-i(\phi_d+\psi)}]\Sigma(\omega)x_{ZPF}}{\hbar|G|D(\omega)}\left[\hat{F}_{ex}(\omega)-\frac{\hbar|G|^2}{\Sigma(\omega)x_{ZPF}}\hat{X}_0(\omega)\right]
\end{align}
where $\hat{F}_{ex}$ is the external force signal, and 
\begin{align}\label{23}
D(\omega)=\chi_c^{-1}(\omega){\chi_c^*}^{-1}(-\omega)-(2\Lambda)^2+2\{\Delta-2\Lambda\sin(2\psi+\phi_d)\}\Sigma(\omega).
\end{align}
It is straightforward to calculate $\hat{a}_{out}^\dag(\omega)$ from Eq. (\ref{22}).
The generalized homodyne cavity output field quadrature mentioned in Eq. (\ref{10}) of Sec. \ref{sec:level3} can be expanded as
\begin{align}\label{25}
 \hat{Z}_{out}(\omega,\theta)=&-\frac{\sqrt{\kappa}\Sigma(\omega)x_{ZPF}[\{\kappa\sin(\theta-\psi)+2\Delta\cos(\theta-\psi)-4\Lambda\sin(\theta+\psi+\phi_d)\}-2i\omega\sin(\theta-\psi)]}{\sqrt{2}\hbar|G|D(\omega)}[\hat{F}_{ex}(\omega)+\hat{f}_N(\omega,\theta)],
 \end{align}
 where the added force noise associated with the input fields entering the system is given by
 \begin{align}\label{26}
 \hat{f}_N(\omega,\theta)=&\frac{-\hbar\sqrt{\gamma}|G|^2}{\Sigma(\omega)x_{ZPF}}[\chi_m(\omega)\hat{b}_{in}(\omega)+\chi_m^*(-\omega)\hat{b}_{in}^\dag(\omega)]\nonumber\\
 &+\frac{\{D(\omega)-\kappa{\chi^*}^{-1}(-\omega)\}e^{-i\theta}-\kappa\{i\Sigma(\omega)e^{-2i\psi}+2\Lambda e^{i\phi_d}\}e^{i\theta}}{\sqrt{\kappa}\frac{\Sigma(\omega)}{\hbar|G|}x_{ZPF}[\{\kappa\sin(\theta-\psi)+2\Delta\cos(\theta-\psi)-4\Lambda\sin(\theta+\psi+\phi_d)\}-2i\omega\sin(\theta-\psi)]}\hat{a}_{in}(\omega)\nonumber\\
 &+\frac{\{D(\omega)-\kappa\chi^{-1}(\omega)\}e^{i\theta}+\kappa\{i\Sigma(\omega)e^{2i\psi}-2\Lambda e^{-i\phi_d}\}e^{-i\theta}}{\sqrt{\kappa}\frac{\Sigma(\omega)}{\hbar|G|}x_{ZPF}[\{\kappa\sin(\theta-\psi)+2\Delta\cos(\theta-\psi)-4\Lambda\sin(\theta+\psi+\phi_d)\}-2i\omega\sin(\theta-\psi)]}\hat{a}_{in}^\dag(\omega).
\end{align}
\section{\label{app:level6}Added force noise power spectral density}
\subsection{\label{app:level6a}Intra-cavity squeezing}
We assume that the optical cavity mode is coupled to the vacuum bath, and that the phonon mode of the MO is coupled to a thermal bath at equilibrium temperature $T$. The input-noise correlators of the cavity and MO are \cite{RevModPhys.82.1155} 

\begin{figure*}[b!]
    \centering
    \includegraphics[width=0.8\textwidth]{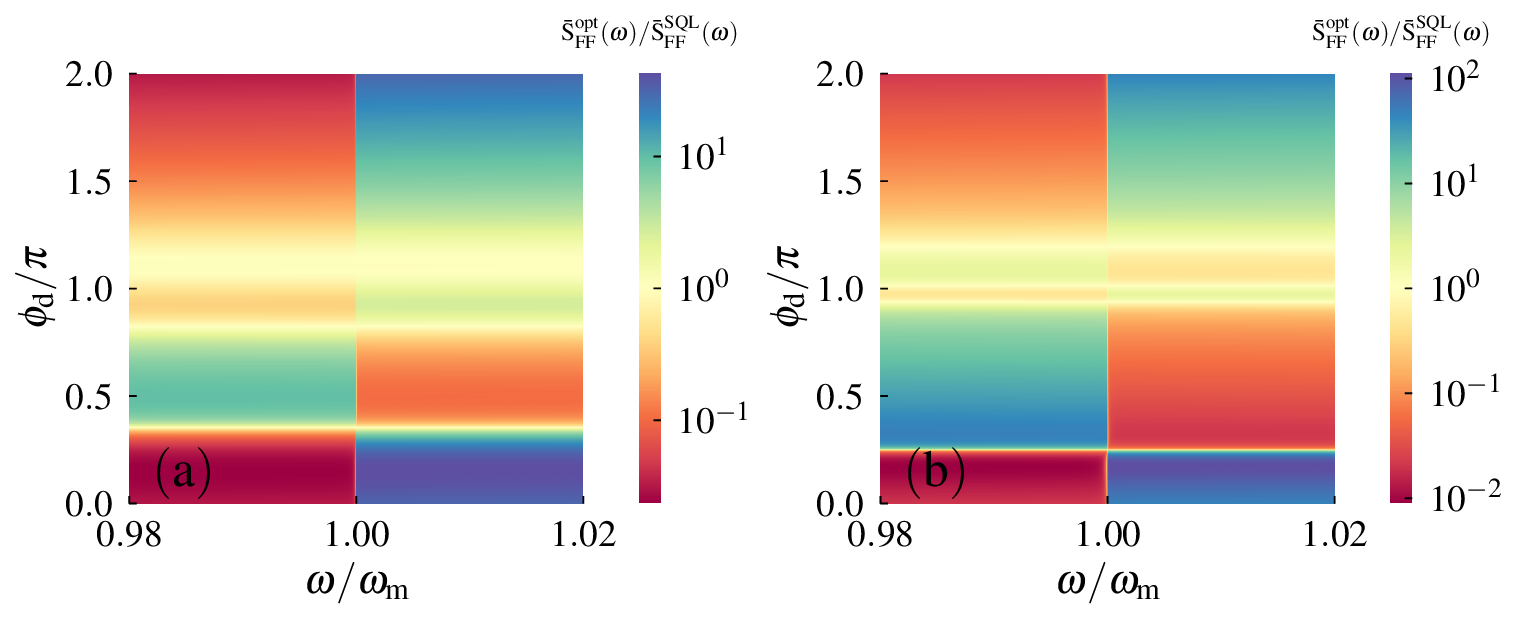}
    \caption{(Color online) The surface plots show the ratio of the optimal added force noise power spectral density to that at SQL as a function of $\phi_d$ and $\omega/\omega_m$ for $\theta=0.9\pi$, (a) $\Lambda=0.15\kappa$, and (b) $\Lambda=0.2\kappa$. The other parameters used for this figure are the same as those used in Fig. \ref{fig:fig2}.}
    \label{fig:fig9}
\end{figure*}

\begin{figure*}[t!]
    \centering
    \includegraphics[width=\textwidth]{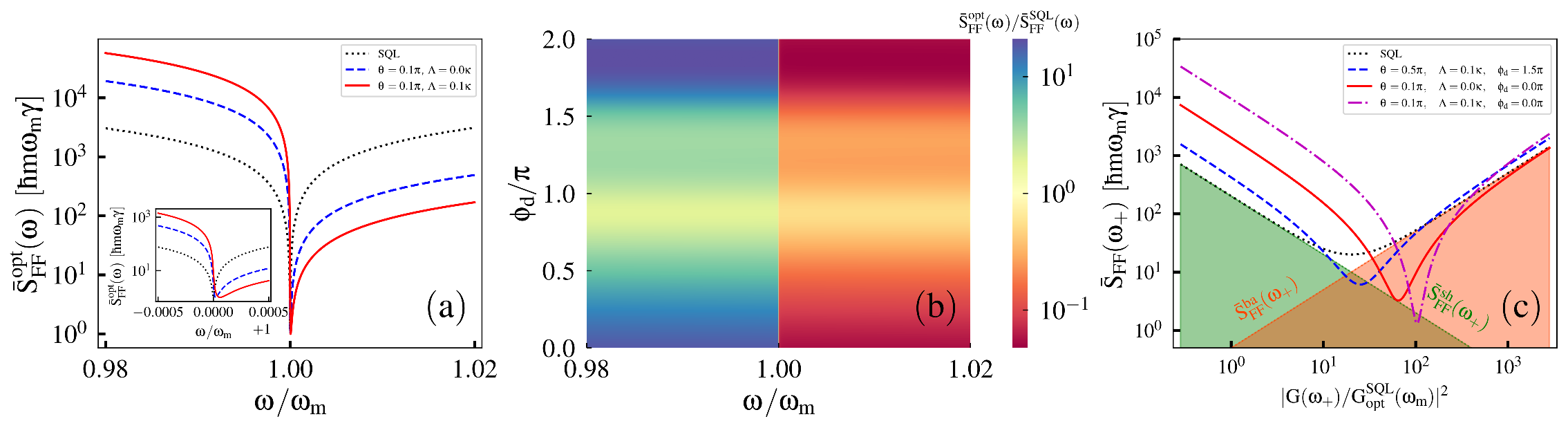}
    \caption{(Color online) (a) The optimal added force noise power spectral density $\bar{S}_{FF}^{opt}(\omega)$ (in the units of $\hbar m\omega_m\gamma$) vs $\omega/\omega_m$ for SQL, $\theta=0.1\pi$, $\phi_d=0$, and different values of $\Lambda$. The inset shows the magnified view in the vicinity of $\omega=\omega_m$. (b) The surface plot shows the ratio of the optimal added force noise power spectral density to that at SQL as a function of $\phi_d$ and $\omega/\omega_m$ for $\theta=0.1\pi$ and $\Lambda=0.1\kappa$. (c) The added force noise power spectral density $\bar{S}_{FF}(\omega_+)$ (in the units of $\hbar m\omega_m\gamma$) vs the normalized probe power $|G(\omega_+)/G_{opt}^{SQL}(\omega_m)|^2$ for a frequency $\omega_+=\omega_m+10\gamma$ with different values of $\theta$, $\Lambda$, and $\phi_d$. The other parameters used for this figure are the same as those used in Fig. \ref{fig:fig2}.}
    \label{fig:fig10}
\end{figure*}

\begin{subequations}
    \begin{align}
        \langle\hat{a}_{in}(\omega)\hat{a}_{in}^\dag(\omega^\prime)\rangle=\delta(\omega+\omega^\prime),&\quad\langle\hat{a}_{in}^\dag(\omega)\hat{a}_{in}(\omega^\prime)\rangle=0,\label{27i}\\
        \langle\hat{b}_{in}(\omega)\hat{b}_{in}^\dag(\omega^\prime)\rangle=[\bar{n}_{th}(\omega)+1]\delta(\omega+\omega^\prime),&\quad\langle\hat{b}_{in}^\dag(\omega)\hat{b}_{in}(\omega^\prime)\rangle=\bar{n}_{th}(\omega)\delta(\omega+\omega^\prime),\label{27ii}\\
        \langle\hat{a}_{in}(\omega)\hat{a}_{in}(\omega^\prime)\rangle=\langle\hat{a}_{in}^\dag(\omega)\hat{a}_{in}^\dag(\omega^\prime)\rangle=&\langle\hat{b}_{in}(\omega)\hat{b}_{in}(\omega^\prime)\rangle=\langle\hat{b}_{in}^\dag(\omega)\hat{b}_{in}^\dag(\omega^\prime)\rangle=0\label{27iii},
    \end{align}
\end{subequations}

where $\bar{n}_{th}(\omega)=[e^{\{\hbar\omega/(k_BT)\}}-1]^{-1}$ is the mean phonon number in equilibrium with the thermal bath. Substituting Eq. (\ref{26}) and its hermitian conjugate (with $\Delta=0$) in Eq.(\ref{11}) of Sec. \ref{sec:level3} and applying the input-noise correlators in Eqs. (\ref{27i}), (\ref{27ii}), and (\ref{27iii}), one can derive the added force noise power spectral density
\begin{align}\label{28}
\bar{S}_{FF}(\omega,\theta)=\bar{S}_{FF}^{th}(\omega,\theta)+\bar{S}_{FF}^{sh}(\omega,\theta)+\bar{S}_{FF}^{ba}(\omega,\theta)+\bar{S}_{FF}^{cor}(\omega,\theta).
\end{align}
The contributions from thermal noise from the bath of mechanical phonons, shot noise from the cavity input, backaction noise from the optomechanical radiation pressure interaction, and the correlation of shot noise and backaction noise are respectively given by
\begin{subequations}\label{29}
    \begin{align}
        \bar{S}_{FF}^{th}(\omega,\theta)=&\frac{\hbar^2\gamma|G|^4[|\chi_m(\omega)|^2+|\chi_m(-\omega)|^2][\bar{n}_{th}(\omega)+1/2]}{|\Sigma(\omega)|^2x_{ZPF}^2},\label{29i}\\
        \bar{S}_{FF}^{sh}(\omega,\theta)=&\frac{A^2+B^2+2AB\cos(2\theta+\phi_d)}{\kappa x_{ZPF}^2\frac{|\Sigma(\omega)|^2}{\hbar^2|G|^2}[\{\kappa\sin(\theta-\psi)-4\Lambda\sin(\theta+\psi+\phi_d)\}^2+\{2\omega\sin(\theta-\psi)\}^2]},\label{29ii}\\
        \bar{S}_{FF}^{ba}(\omega,\theta)=&\frac{\hbar^2|G|^2[C^2+D^2-2CD\cos(\theta-\psi)]}{\kappa x_{ZPF}^2[\{\kappa\sin(\theta-\psi)-4\Lambda\sin(\theta+\psi+\phi_d)\}^2+\{2\omega\sin(\theta-\psi)\}^2]},\label{29iii}\\
        \bar{S}_{FF}^{cor}(\omega,\theta)=&\frac{-2Re[\Sigma(\omega)][C\{A\cos(\theta+\psi+\phi_d)+B\cos(\theta-\psi)\}-D\{A\cos(2\theta+\phi_d)+B\}]}{\kappa x_{ZPF}^2\frac{|\Sigma(\omega)|^2}{\hbar^2|G|^2}[\{\kappa\sin(\theta-\psi)-4\Lambda\sin(\theta+\psi+\phi_d)\}^2+\{2\omega\sin(\theta-\psi)\}^2]},\label{29iv}
    \end{align}
\end{subequations}
where $A=2\kappa\Lambda$, $B=\omega^2+\kappa^2/4+(2\Lambda)^2$, $C=2\kappa\sin(\theta-\psi)$, and $D=4\Lambda\sin(2\psi+\phi_d)$. We note that $\psi$ is a function of $\Lambda$ and $\phi_d$ for a fixed $\kappa$ and $\Delta=0$ as given by Eq. (\ref{5b}) of Sec. \ref{sec:level2}. Using the expression of the mechanical self-energy $\Sigma(\omega)$ given in Appendix \ref{app:level5}, one can show that the shot noise  (\ref{29ii}) is inversely proportional to $|G|^2$ and the backaction noise (\ref{29iii}) is directly proportional to $|G|^2$. However, the thermal noise (\ref{29i}) and the shot noise-backaction noise quantum correlation (\ref{29iv}) do not explicitly depend on $|G|^2$. We have neglected other technical noises for simplicity.
\subsection{\label{app:level6b}Injected external squeezing}
In case of the external injection of squeezed vacuum into the cavity in the absence of ICS, the input noise correlators of the MO remain the same as Eqs. (\ref{27ii}), and (\ref{27iii}). However, the input-noise correlators of the cavity are  
\begin{subequations}\label{30}
    \begin{align}
        \langle\hat{a}_{in}^\dag(\omega)\hat{a}_{in}(\omega^\prime)\rangle=N\delta(\omega+\omega^\prime),&\quad\langle\hat{a}_{in}(\omega)\hat{a}_{in}^\dag(\omega^\prime)\rangle=(N+1)\delta(\omega+\omega^\prime),\\
        \langle\hat{a}_{in}(\omega)\hat{a}_{in}(\omega^\prime)\rangle=M\delta(\omega+\omega^\prime),&\quad\langle\hat{a}_{in}^\dag(\omega)\hat{a}_{in}^\dag(\omega^\prime)\rangle=M^*\delta(\omega+\omega^\prime)
    \end{align}
\end{subequations}
where $M=e^{i\phi_e}\sinh(r_e)\cosh(r_e)$ and $N=\sinh^2(r_e)$, with $r_e$ and $\phi_e$ denoting the squeezing parameter and squeezing angle, respectively \cite{alma990014403420203701, PhysRevA.94.051801}. Since the thermal phonon bath of the MO remains unchanged, $\bar{S}_{FF}^{th}(\omega,\theta)$ is the same as Eq. (\ref{29i}). However, the other noise contributions are given below.
\begin{figure*}[t!]
    \centering
    \includegraphics[width=\textwidth]{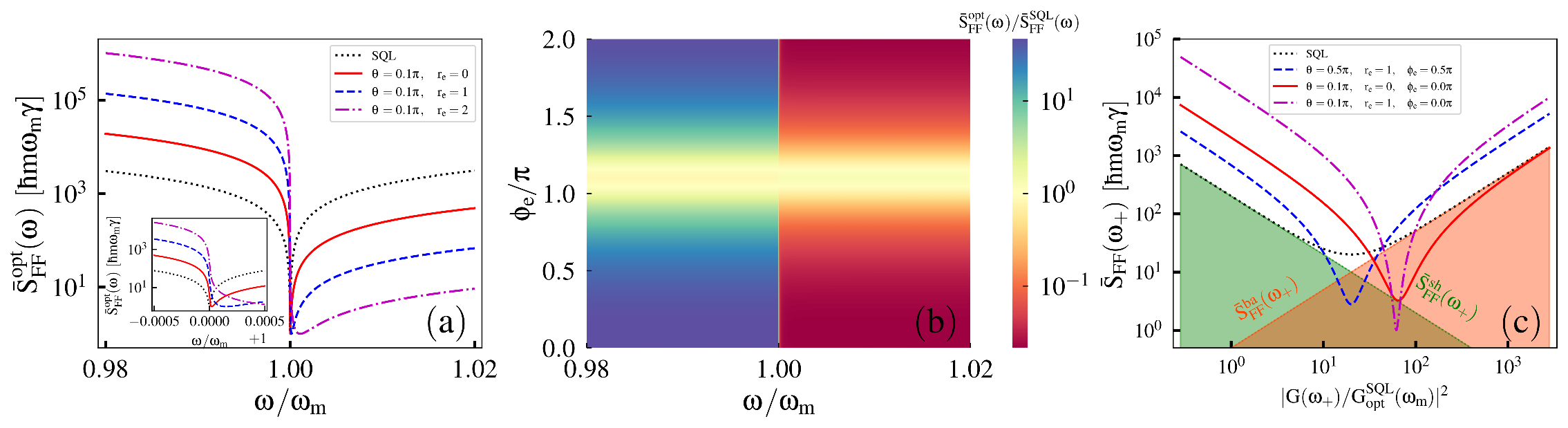}
    \caption{(Color online) (a) The optimal added force noise power spectral density $\bar{S}_{FF}^{opt}(\omega)$ (in the units of $\hbar m\omega_m\gamma$) vs $\omega/\omega_m$ for SQL, $\theta=0.1\pi$, $\phi_e=0$, and different values of $r_e$. The inset shows the magnified view in the vicinity of $\omega=\omega_m$. (b) The surface plot shows the ratio of the optimal added force noise power spectral density to that at SQL as a function of $\phi_e$ and $\omega/\omega_m$ for $\theta=0.1\pi$ and $r_e=1$. (c) The added force noise power spectral density $\bar{S}_{FF}(\omega_+)$ (in the units of $\hbar m\omega_m\gamma$) vs the normalized probe power $|G(\omega_+)/G_{opt}^{SQL}(\omega_m)|^2$ for a frequency $\omega_+=\omega_m+10\gamma$ with different values of $\theta$, $r_e$, and $\phi_e$. The other parameters used for this figure are the same as those used in Fig. \ref{fig:fig2}.}
    \label{fig:fig11}
\end{figure*}

\begin{subequations}\label{31}
    \begin{align}
        \bar{S}_{FF}^{sh}(\omega,\theta)=&\frac{(\omega^2+\kappa^2/4)[\cos(2\theta-\phi_e)\sinh(2r_e)+\cosh(2r_e)]}{\kappa x_{ZPF}^2\frac{|\Sigma(\omega)|^2}{\hbar^2|G|^2}[4\sin^2(\theta)]},\label{31i}\\
        \bar{S}_{FF}^{ba}(\omega,\theta)=&\frac{\hbar^2|G|^2[2\kappa\sin(\theta)]^2[\cos(\phi_e)\sinh(2r_e)+\cosh(2r_e)]}{\kappa x_{ZPF}^2[4(\omega^2+\kappa^2/4)]\sin^2(\theta)},\label{31ii}\\
        \bar{S}_{FF}^{cor}(\omega,\theta)=&\frac{-2Re[\Sigma(\omega)][2\kappa\sin(\theta)][\cos(\theta-\phi_e)\sinh(2r_e)+\cosh(2r_e)\cos(\theta)]}{\kappa x_{ZPF}^2\frac{|\Sigma(\omega)|^2}{\hbar^2|G|^2}[4\sin^2(\theta)]}.\label{31iii}
    \end{align}
\end{subequations}
The dependence of different noise sources on $|G|^2$ for IES is the same as that for ICS, as discussed in Appendix \ref{app:level6a}.
\end{widetext}
\section{\label{app:level7}Stability analysis}
First, we study the stability of our model system by examining the behavior of the steady-state intra-cavity photon number $|\alpha|^2$. The exact solution of $|\alpha|^2$ for $\Delta_c=0$ can be obtained from Eq. (\ref{5a}) of Sec. \ref{sec:level2}, by solving 
\begin{align}\label{32}
a_0|\alpha|^6+a_1|\alpha|^4+a_2|\alpha|^2+a_3=0,
\end{align}
where the coefficients are
\begin{align}\label{33}
        a_0&=\beta^{\prime^2},\quad a_1=-\Omega^2\beta^{\prime^2},\nonumber\\
        a_2&=[\kappa^2/4-(2\Lambda)^2]+4\Lambda\Omega^2\beta^\prime\sin(\phi_d),\nonumber\\
        a_3&=-\Omega^2[\kappa^2/4+2\kappa\Lambda\cos(\phi_d)+(2\Lambda)^2],
\end{align}

\begin{figure}[b!]
    \centering
    \includegraphics[width=.4\textwidth]{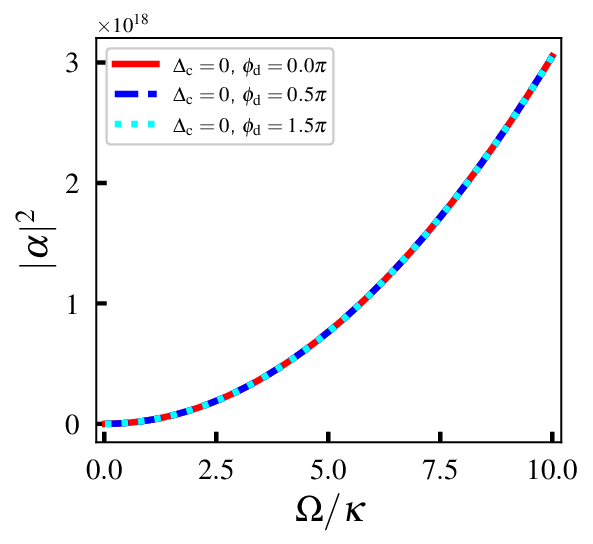}
    \caption{(Color online) The steady-state photon number $|\alpha|^2$ vs the normalized driving strength $\Omega/\kappa$ of the cavity probe field for $\Delta_c=0$. The parameters used for this plot are $\Lambda=0.1\kappa$ and the other parameters are the same as those mentioned in Fig. \ref{fig:fig2}.}
    \label{fig:fig12}
\end{figure}

and $\beta^\prime=\frac{2g^2\omega_m}{\omega_m^2+\gamma^2/4}$. Fig \ref{fig:fig12} delineates the monostable behaviour of $|\alpha|^2$ over a wide range of $\Omega$ for the given system parameters. We observe that $|\alpha|^2$ does not vary with a change in the two-photon drive phase angle $\phi_d$.

The Routh-Hurwitz criterion (RHC) can be applied to constrain the system parameters and ensure the system stability \cite{PhysRevA.35.5288}. The linearized equation of motion of our model system can be written as 
\begin{figure}[b!]
    \centering
    \includegraphics[width=0.49\textwidth]{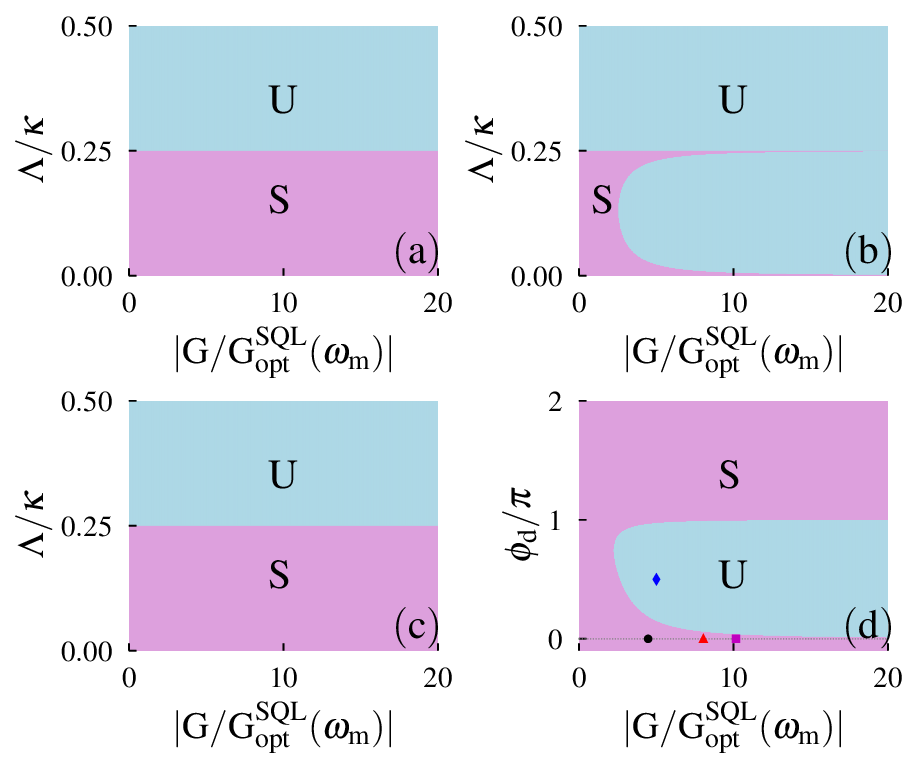}
    \caption{(Color online)  The stable `$\mathrm{S}$' (plum) and unstable `$\mathrm{U}$' (lightblue) regions are shown as a function of the normalized two-photon drive amplitude $\Lambda/\kappa$ and the normalized linearized optomechanical coupling $|G/G_{opt}^{SQL}(\omega_m)|$ for (a) $\phi_d=0$, (b) $\phi_d=0.5\pi$, and (c) $\phi_d=1.5\pi$. (d) Stability as a function of the normalized two-photon drive phase $\phi_d/\pi$ and $|G/G_{opt}^{SQL}(\omega_m)|$ for $\Lambda=0.1\kappa$. The other parameters used in this figure are the same as those used in Fig. \ref{fig:fig2}.}
    \label{fig:fig13}
\end{figure}
\begin{align}\label{34}
\dot{\mathcal{R}}(t)=\mathcal{A}\mathcal{R}(t)+\mathcal{B}\mathcal{R}_{in}(t),
\end{align}
where $\mathcal{R}(t)=[\hat{a}(t),\hat{a}^\dag(t),\hat{b}(t),\hat{b}^\dag(t)]^T$ represents the operator vector, $\mathcal{B}=[\sqrt{\kappa},\sqrt{\kappa},\sqrt{\gamma},\sqrt{\gamma}]$, $\mathcal{R}_{in}(t)=[\hat{a}_{in}(t),\hat{a}_{in}^\dag(t),\hat{b}_{in}(t),\hat{b}_{in}^\dag(t)]^T$, and
\begin{align}\label{35}
\mathcal{A}=\begin{pmatrix}
    -(i\Delta+\frac{\kappa}{2}) & 2\Lambda e^{-i\phi_d} & iG & iG\\
    2\Lambda e^{i\phi_d} & (i\Delta-\frac{\kappa}{2}) & -iG^*  & -iG^*\\
    iG^* & iG & -(i\omega_m+\frac{\gamma}{2}) & 0\\
    -iG^* & -iG & 0 & (i\omega_m-\frac{\gamma}{2})
\end{pmatrix}.
\end{align}
For $\Delta_c=0$, we assume $\Delta\approx0$ or $\pm i\Delta-\frac{\kappa}{2}\approx-\frac{\kappa}{2}$ as $g(\beta+\beta^*)\ll\frac{\kappa}{2}$ for the given system parameters. The stability of the system is determined by the eigenvalues of the matrix $\mathcal{A}$. The system is stable if all the eigenvalues have negative real parts, so that the oscillating parts of the steady-state solutions of Eq. (\ref{34}) decay rapidly. 

Fig \ref{fig:fig13} (a) and (c) depict the stability of the system (plum colour) for $\Lambda<0.25\kappa$ with $\phi_d=0$ $(2n\pi)$ and $\phi_d=1.5\pi$, respectively. However, the system becomes unstable (light-blue colour) for larger values of $|G|$ for $\Lambda<0.25\kappa$ with $\phi_d=0.5\pi$ as shown in Fig. \ref{fig:fig13} (b). Fig. \ref{fig:fig13} (d) indicates that the system remains in the stable region for $\Lambda=0.1\kappa$ and $\phi_d>\pi$, while it becomes unstable for $0<\phi_d<\pi$ with larger values of $|G|$. For $\Delta\approx0$ and $\Lambda=\phi_d=0$, the system remains stable for arbitrary values of the other system parameters, although we have not shown in Fig \ref{fig:fig13}.

\bibliographystyle{apsrev4-2}
\bibliography{ref}
\end{document}